\documentclass[12pt,a4paper,dvips]{article}
\usepackage{a4p}
\usepackage{cite,mcite}
\usepackage{graphicx}
\usepackage{epsfig}
\usepackage{physics}
\usepackage{l3_titlePH,ifthen}

\journalname{Phys. Lett. B}

\date{November 1, 2004}

\preprint{2004-055}

\newlength{\capindent}
\newlength{\capwidth}
\newlength{\figwidth}
\newcommand{\icaption}[2][!*!,!]{\hspace*{\capindent}%
  \begin{minipage}{\capwidth}
    \ifthenelse{\equal{#1}{!*!,!}}%
      {\caption{#2}}%
      {\caption[#1]{#2}}
  \end{minipage}}

\newcommand{\EE}{\ensuremath{\mathrm{e^+ e^-}}}

\newcommand{\MM}{\ensuremath{\mathrm{\mu^+ \mu^-}}}

\newcommand{\TT}{\ensuremath{\mathrm{\tau^+\tau^-}}}

\begin{document}
%

\begin{titlepage}

\title{\boldmath Search for an Invisibly-Decaying Higgs Boson at LEP}

\author{L3 Collaboration}

\begin{abstract}

A search for a Higgs boson produced in $\rm e^+e^-$ collisions in
association with a Z boson and decaying into invisible particles is
performed. Data collected at LEP with the L3 detector at
centre-of-mass energies from $189\gev$ to $209\gev$ are used,
corresponding to an integrated luminosity of 0.63~fb$^{-1}$. Events
with hadrons, electrons or muons with visible masses compatible with a
Z boson and missing energy and momentum are selected. They are
consistent with the Standard Model expectations.  A lower limit of
$112.3\gev{}$ is set at $95\%$ confidence level on the mass of the
invisibly-decaying Higgs boson in the hypothesis that its production
cross section equals that of the Standard Model Higgs boson. Relaxing
this hypothesis, upper limits on the production cross section are
derived.

\end{abstract}

\submitted
\end{titlepage}

%
\section{Introduction}
%

The Standard Model of the electroweak
interactions~\cite{standard_model} relies on the Higgs
mechanism~\cite{higgs_mech} to explain the observed masses of the
elementary particles. A consequence of this mechanism is the existence
of an additional particle, the Higgs boson. Direct searches at the LEP
$\epem$ collider for the Standard Model Higgs boson, H, produced in
the Higgs-strahlung process $\epem\ra\rm{H}\Zo$ did not observe a
significant excess of events over the Standard Model
expectations~\cite{higgs,aleph_higgs,lnq}. These searches are based on
the hypothesis that the Higgs boson mainly decays into b
quarks. Searches in which this hypothesis is relaxed and the Higgs
boson is allowed to decay into a generic hadronic final state also
yield negative results~\cite{flav_ind}. In addition, no signs of the
Higgs boson were found in cases in which anomalous couplings would
affect its production and decay mechanisms~\cite{ano_h}.

However, a Higgs boson which decays into stable weakly-interacting
particles would have escaped detection in all these searches. Such
possibility has been extensively proposed in literature for the case
of Higgs bosons decaying into the lightest supersymmetric
particles~\cite{hinv_th},  fourth-generation
neutrinos~\cite{neutrinos},  neutrinos in the context of theories
with extra space dimensions~\cite{martin}, 
majorons~\cite{majoron,martin} or into a general scalar gauge singlet
added to the Standard Model~\cite{general}.

This Letter describes the search for an invisibly-decaying Higgs
boson, h, produced through the Higgs-strahlung process $\rm\epem\ra h
Z$. Decays of the Z boson into hadrons, electron pairs and muon pairs
are considered and analyses are devised to select events with hadrons
or leptons and missing energy and momentum. Data collected by the L3
detector~\cite{l3det} at LEP at centre-of-mass energies $\sqrt{s} =
189 -209\GeV$ are analysed. They correspond to a total integrated
luminosity of 0.63~fb$^{-1}$, as detailed in Table~\ref{tab:data}.

The results presented in this Letter supersede those of previous L3
studies~\cite{Acciarri:1997hinv,L3old}, as the complete high-luminosity
and high-energy data sample is investigated and the previously
published data collected at $\sqrt{s} = 189\GeV$~\cite{L3old} are
re-analysed with improved procedures.  Similar searches were also
performed by other LEP collaborations~\cite{aleph_higgs,otherLep}.

%
\section{Event simulation}
%

To optimise the selection criteria and determine the efficiency to
detect a possible signal, samples of Higgs-boson events are generated
using the PYTHIA Monte Carlo program \cite{pythia} for masses of the
Higgs boson, $m_{\rm h}$, between $50\gev{}$ and $120\gev$ in steps between
$5\gev$ and $10\gev$.

The following Monte Carlo programs are used to model Standard Model
processes: KK2f~\cite{kk2f} for $\rm\epem\ra q \bar{q}$,
$\rm\epem\ra\mu^+\mu^-$ and $\rm\epem\ra\tau^+\tau^-$,
BHWIDE~\cite{bhwide} for Bhabha scattering, and PHOJET~\cite{phojet}
and DIAG36~\cite{diag36} for hadron and lepton production in
two-photon collisions, respectively. Four-fermion final states
relevant for the analysis of events with hadrons and missing energy
are generated with PYTHIA for Z-boson
pair-production and the $\rm\epem\ra Z \epem$ process and with KORALW~\cite{koralw} for W-boson pair-production,
with the exception of the $\rm\epem\ra W e \nu\ra qq e \nu$ process,
modelled with EXCALIBUR~\cite{excalibur}.  All
four-fermion processes with charged leptons and neutrinos in the final
states, relevant for the analysis of events with leptons and missing
energy, are generated with KandY~\cite{KandY}.

For each centre-of-mass energy, the number of simulated background
events corresponds to at least 50 times the number of expected events,
up to a maximum of 7.5 million KandY events, except for two-photon
interactions and Bhabha scattering for which twice and seven times the
collected luminosity is simulated, respectively.

The L3 detector response is simulated using the GEANT
program~\cite{my_geant}, which takes into account the effects of
energy loss, multiple scattering and showering in the detector.
Time-dependent inefficiencies of the different subdetectors, as
monitored during the data-taking period, are taken into account in the
simulation procedure.

%
\section{Selection of events with hadrons and missing energy}
%

A preselection identifies events compatible with the production of a
heavy invisible particle and a Z boson decaying into hadrons. High
multiplicity events are retained if their visible energy, $E_{vis}$,
satisfies $0.3<E_{vis}/\sqrt{s}<0.65$ and have a visible mass between
$60\GeV$ and $115\GeV$. No identified leptons or photons of energy
above $10\GeV$ are allowed in these events. To suppress the large
background from hadron production in two-photon collisions and events
from the $\epem\ra\rm q\bar{q}\gamma$ process with a high-energy and
low polar-angle photon, the missing momentum of the event is required
to point inside the detector: its polar angle with respect to the beam
axis, $\theta_{miss}$, must satisfy $|\cos{\theta_{miss}}| < 0.9$. In
addition, the event is reconstructed into two jets by means of the
DURHAM algorithm~\cite{DURHAM} and the angle between the jets is required to be smaller than $175^{\circ}$. Events 
with large energy deposits in the low-angle calorimeters are also
rejected. After the preselection, 779 events are selected in data
while 772 events are expected from Standard Model processes, as
detailed in Table~\ref{tab:hadrons}. The signal efficiencies depend on
$m_{\rm h}$, and vary from 52\% up to 59\%. Up to 90\% of the background
comes from four-fermion processes and 10\% from fermion-pair production.

Two selections are devised in order to retain high efficiency for
light and heavy Higgs bosons. The ``light-Higgs selection'' is applied
to events where the relativistic velocity of the reconstructed hadron
system, $\beta$, satisfies $\beta>0.4$. The ``heavy-Higgs selection''
is applied to the remaining events.

\begin{figure}[p]
  \begin{center}
    \begin{tabular}{cc}
      \mbox{\epsfig{figure=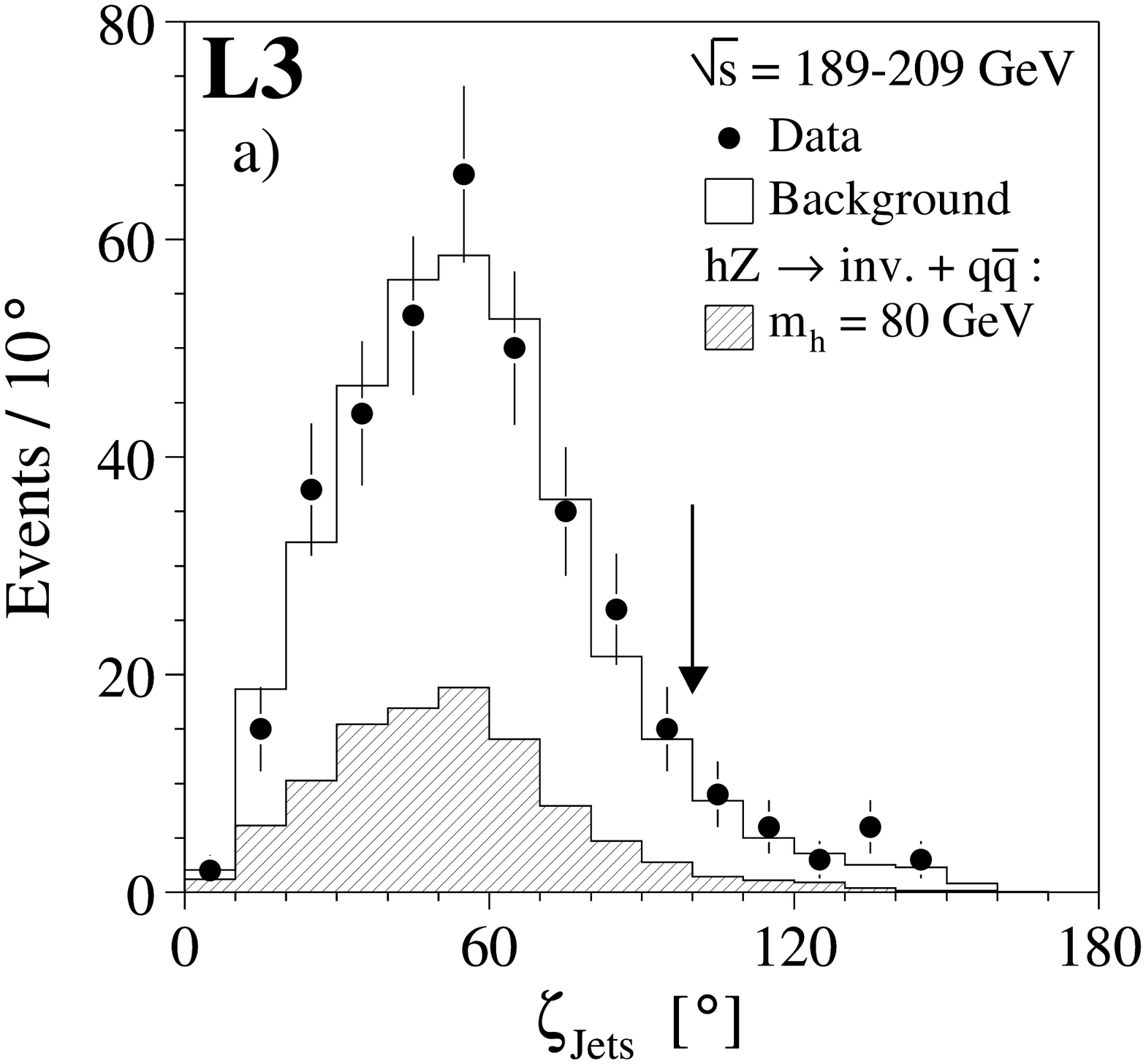,width=0.5\textwidth}}&
      \mbox{\epsfig{figure=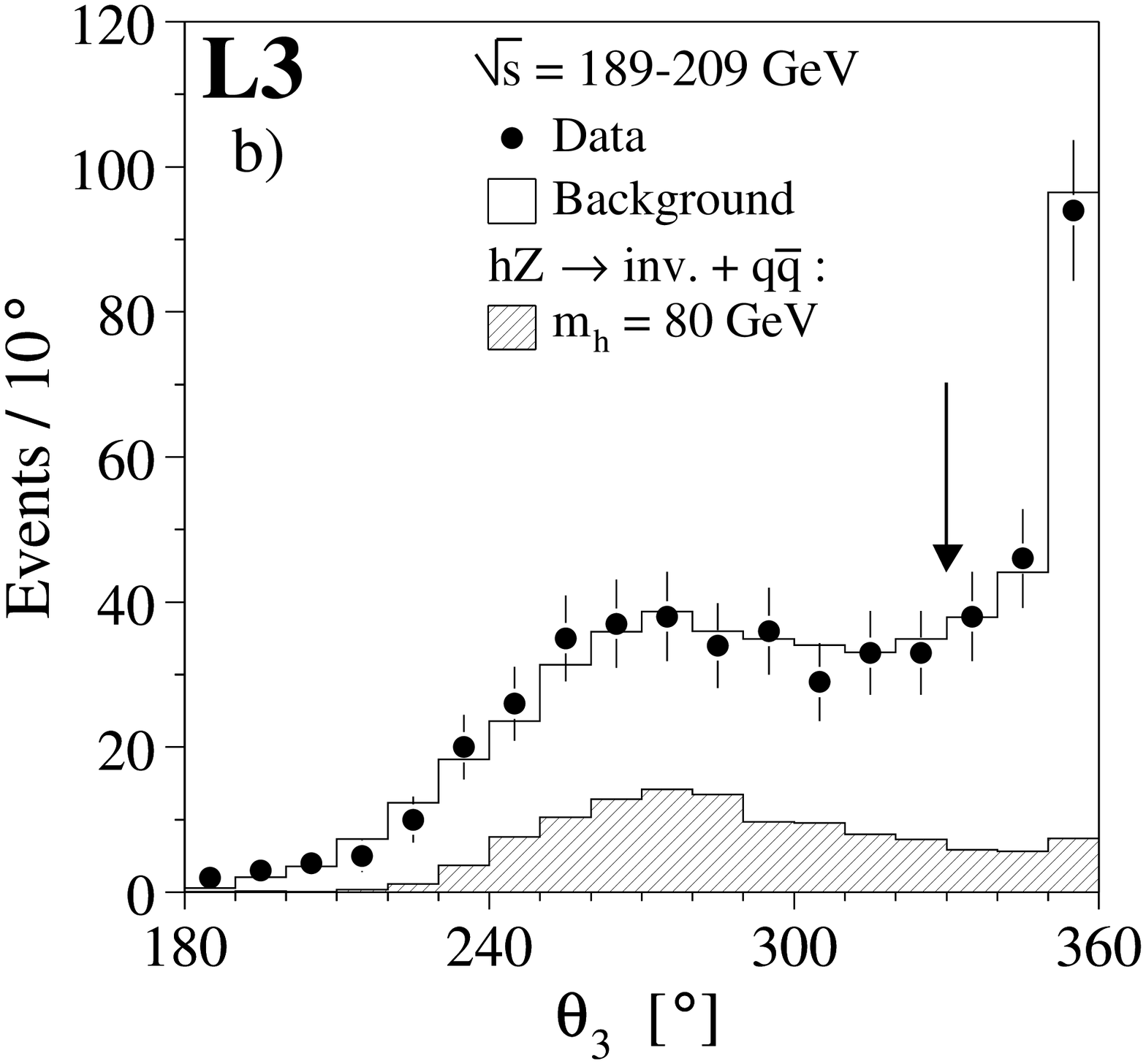,width=0.5\textwidth}}\\
      \mbox{\epsfig{figure=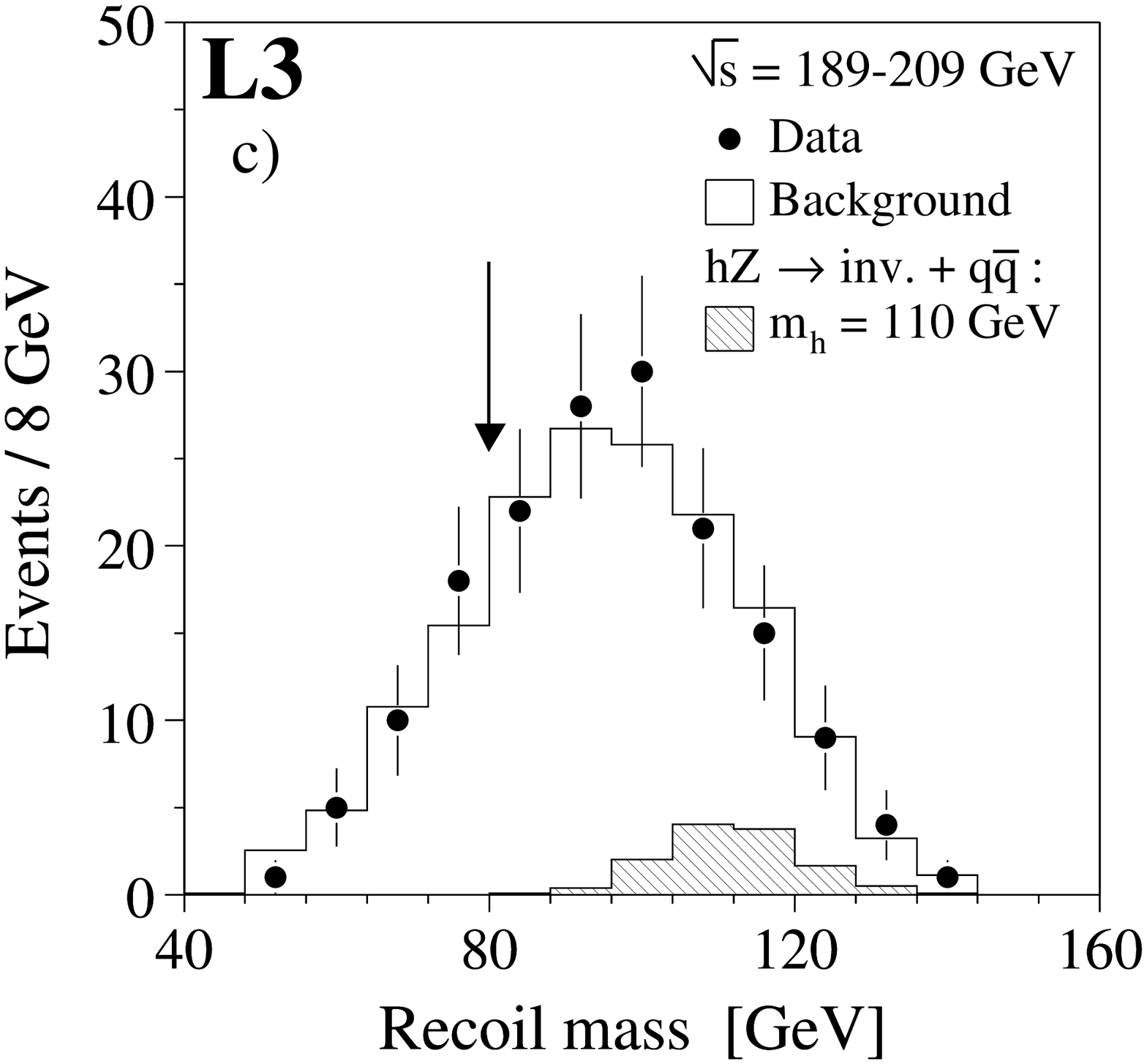,width=0.5\textwidth}}&
      \mbox{\epsfig{figure=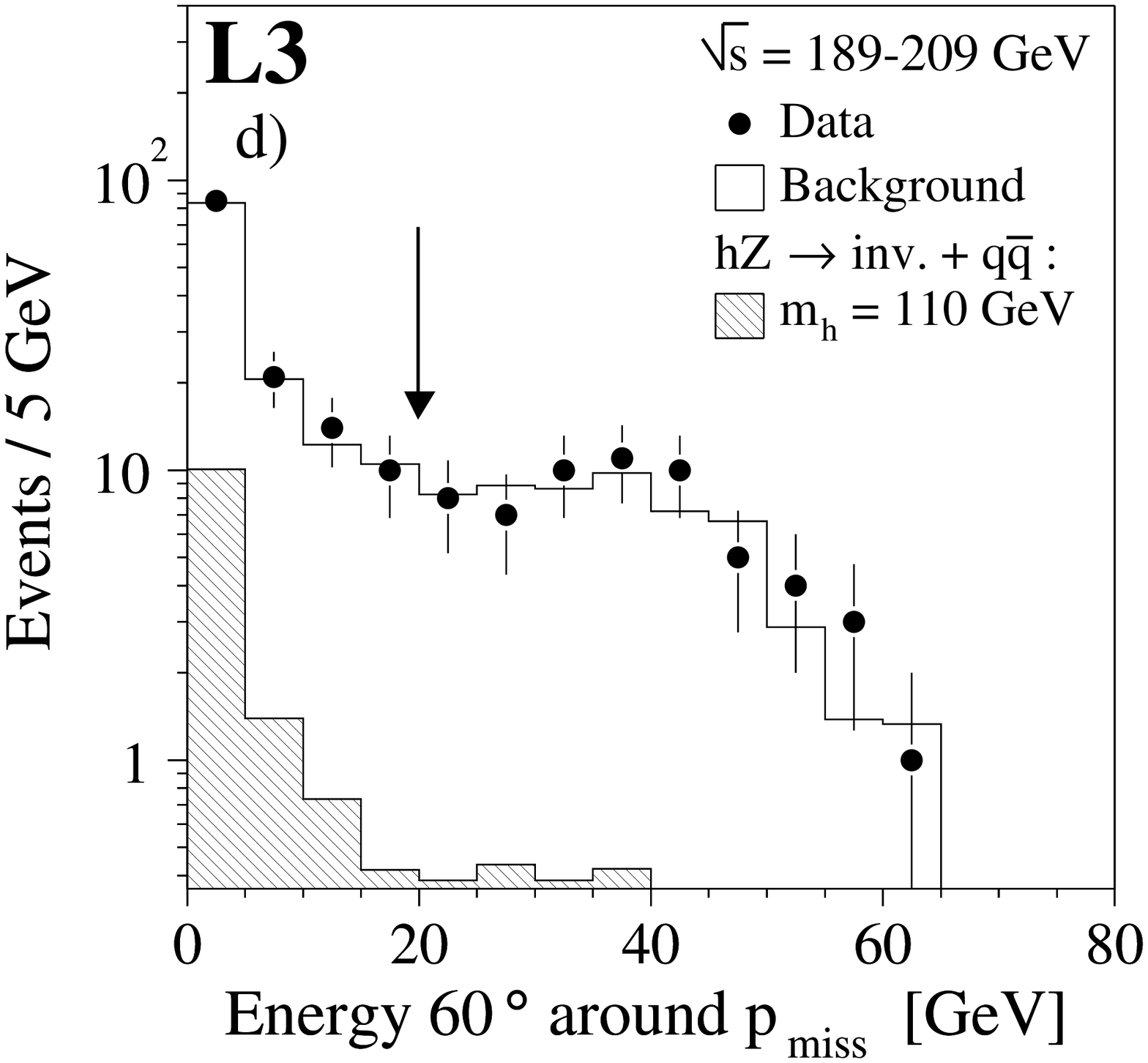,width=0.5\textwidth}}\\
    \end{tabular}
    \icaption[]{\label{fig:1} Distributions of variables used in the
    selection of events with hadrons and missing energy. The dots
    represent the data, the open area the sum of all background
    contributions and the hatched histogram the expectation for a
    signal. The arrows represent the position of the selection criteria.}
  \end{center}
\end{figure}

The dominant background for the light-Higgs selection arises from W
boson pair-production where one of the W bosons decays into hadrons
and the other into leptons and from the $\epem\ra\rm W e \nu$
process. Two additional selection criteria are applied to reduce these
backgrounds: $\zeta_{jet} < 100^\circ$ and $\theta_3 < 330^\circ$,
where $\zeta_{jet}$ is the angle between the jets in the plane
transverse to the beam direction and $\theta_3$ is the sum of the
three inter-jet angles defined if the event is reconstructed into a
three-jet topology with the DURHAM algorithm. The last cut rejects
genuine three-jet events from W-boson pair-production where a W boson
decays into hadrons and the other into tau leptons which decay into
hadrons. Figures~\ref{fig:1}a and~\ref{fig:1}b present the
distributions of $\zeta_{jet}$ and $\theta_3$.

The heavy-Higgs selection enforces the topology of a heavy undetected
particle by means of two cuts against the background from pair
production of either W bosons or fermions. The mass recoiling to the
hadron system is required to be greater than $80 \GeV$ and the energy
deposited in the calorimeters in a $60^\circ$ cone around the
direction of the missing momentum is required to be smaller than
$20\GeV$.  The distributions of these variables are shown in
Figures~\ref{fig:1}c and~\ref{fig:1}d.

The results of the light- and heavy-Higgs selections are
presented in Table~\ref{tab:hadrons}, while Table~\ref{tab:effi} lists
the signal efficiencies. In total, 475 events are selected in data and
474 are expected from Standard Model processes, dominated by four
fermion final-states.  Figures~\ref{fig:2}a--d present the distributions
of the visible mass of the hadronic system and of the mass recoiling
to the hadronic system, for the light- and heavy-Higgs selections. No
indication for an excess of events in the signal regions is observed.

The sensitivity to a possible Higgs signal is enhanced by building a
discriminating variable for each of the two analyses. This variable
combines~\cite{MSSMHiggs} information from the visible and recoil
masses, as well as the jet widths and the parameter $y_{23}$ of the
DURHAM algorithm for which three jets are reconstructed in a two-jet
event. Figures~\ref{fig:2}e and~\ref{fig:2}f present the distributions
of the discriminating variable for events selected by the light- and
heavy-Higgs selections, respectively. A good agreement between the
observations and the Standard Model predictions is observed.

%
\section{Selection of events with leptons and missing energy}
%

The selection of events possibly originating from an
invisibly-decaying Higgs boson and a Z boson decaying into leptons
proceeds from the L3 analysis of W boson pair-production where either
both W bosons decay into an electron and a neutrino, or both decay
into a muon and a neutrino~\cite{L3WW}. Events with an electron or a
muon pair are selected if the least and most energetic leptons have
energies above $5\GeV$ and $25\GeV$, respectively. The angle of the
leptons with respect to the beam direction, $\theta$, must satisfy
$|\cos\theta|<0.96$. In the case of electrons,
to reduce the background from the forward-peaked
Bhabha scattering, at least one of the
electrons must satisfy $|\cos\theta|<0.92$. To suppress background
from fermion pair-production and cosmic rays, the angle between the
two leptons in the plane transverse to the beam direction,
$\zeta_\ell$, must satisfy $\zeta_\ell < 172^\circ$. Residual
background from cosmic rays is rejected by requiring the leptons to
have a signal in the scintillator time-of-flight counters compatible
with the beam crossing. Finally, the presence of undetected particles
is enforced by requiring the event momentum transverse to the beam
direction, $p_t$, to be greater than $8 \GeV$. 

\begin{figure}[p]
  \begin{center}
    \begin{tabular}{cc}
      \mbox{\epsfig{figure=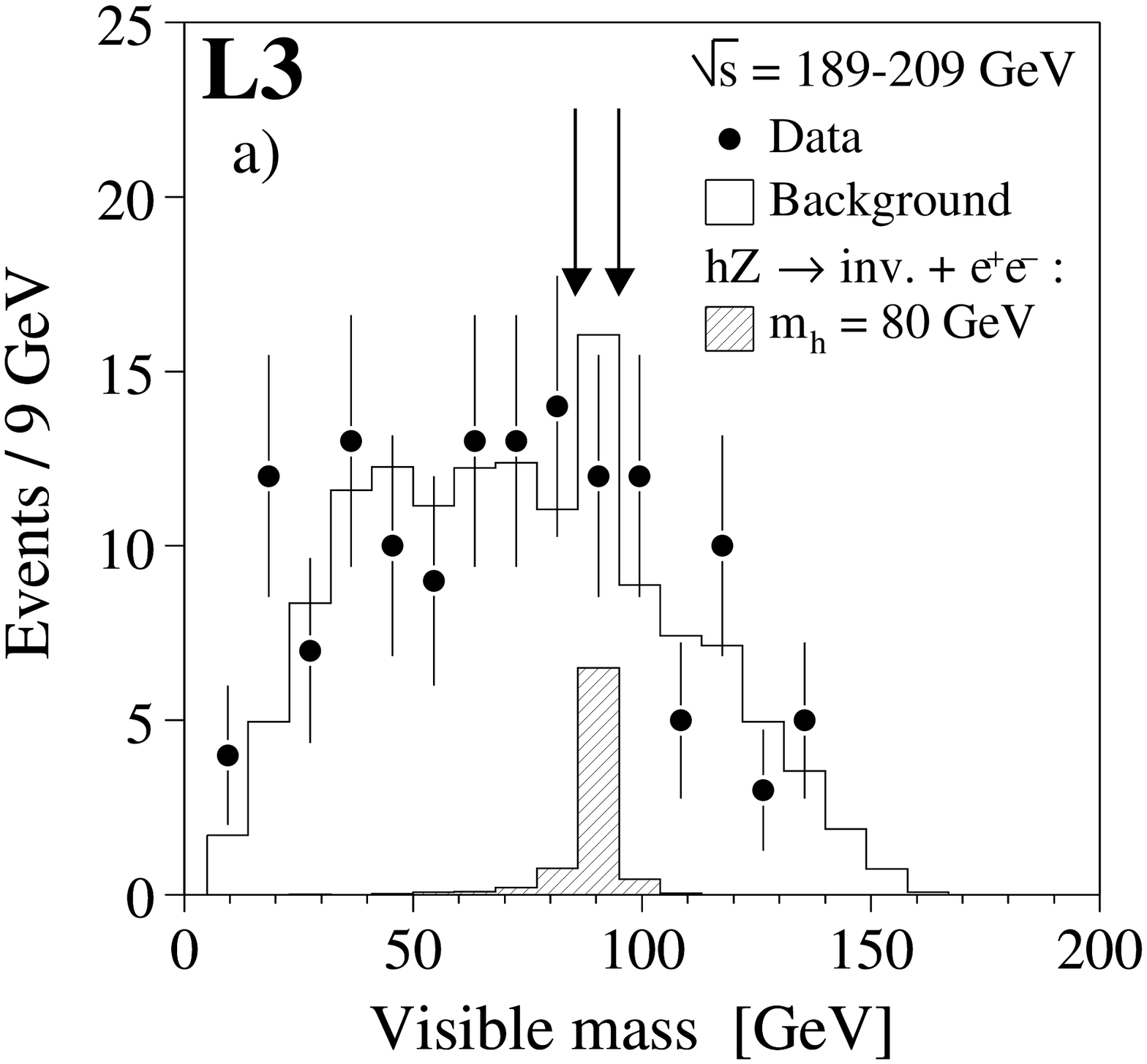,width=0.45\textwidth}}&
      \mbox{\epsfig{figure=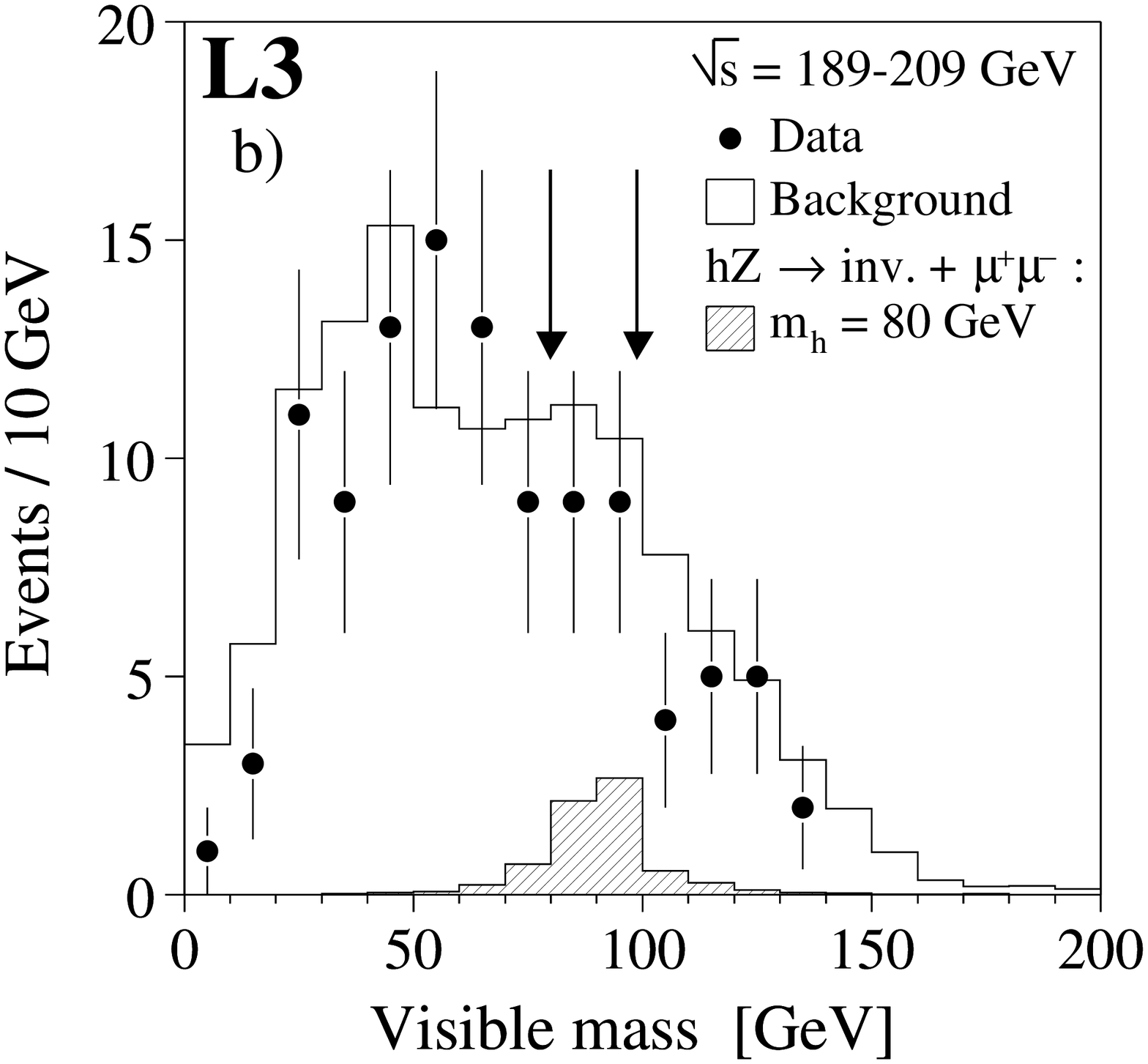,width=0.45\textwidth}}\\
      \mbox{\epsfig{figure=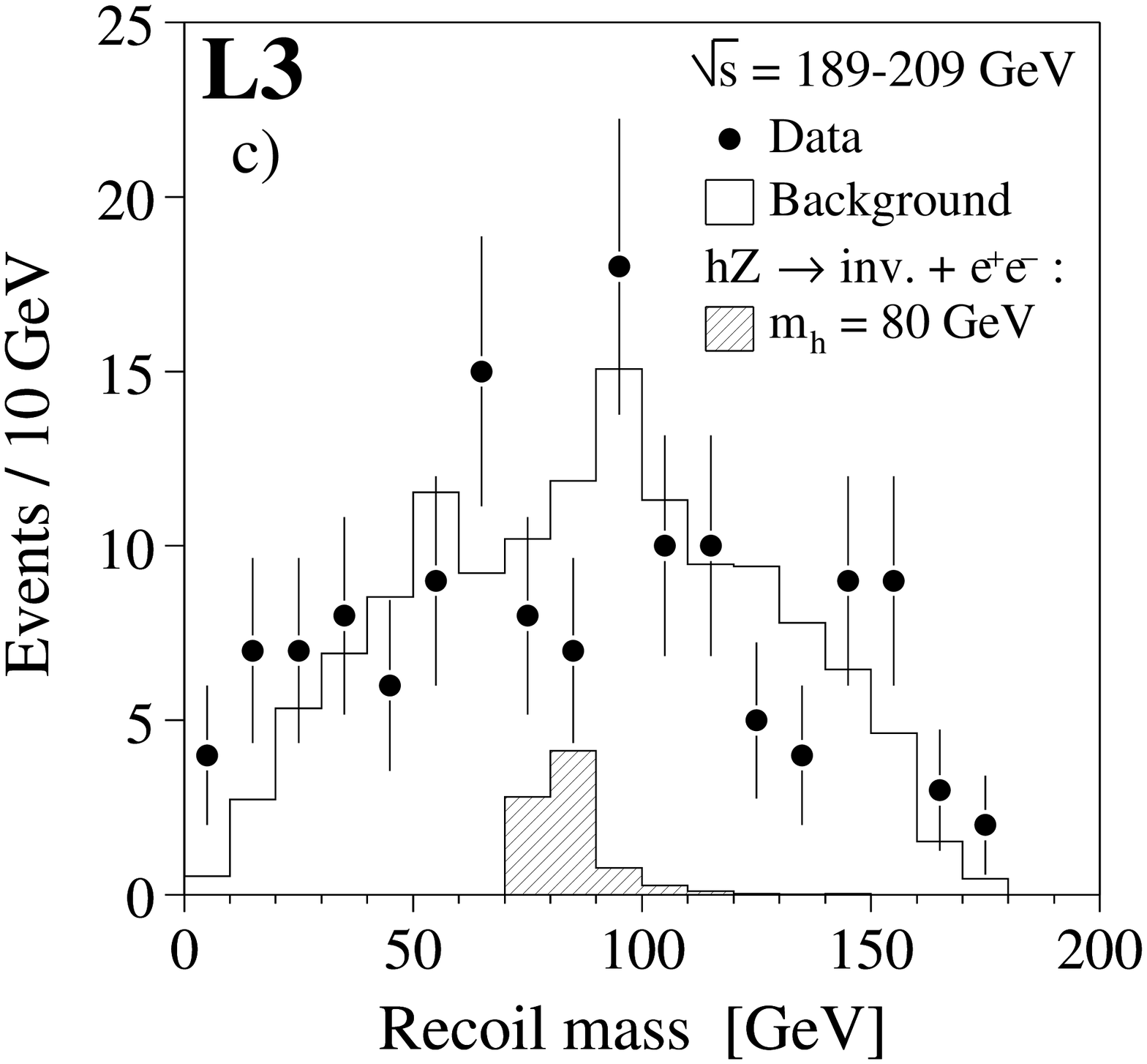,width=0.45\textwidth}}&
      \mbox{\epsfig{figure=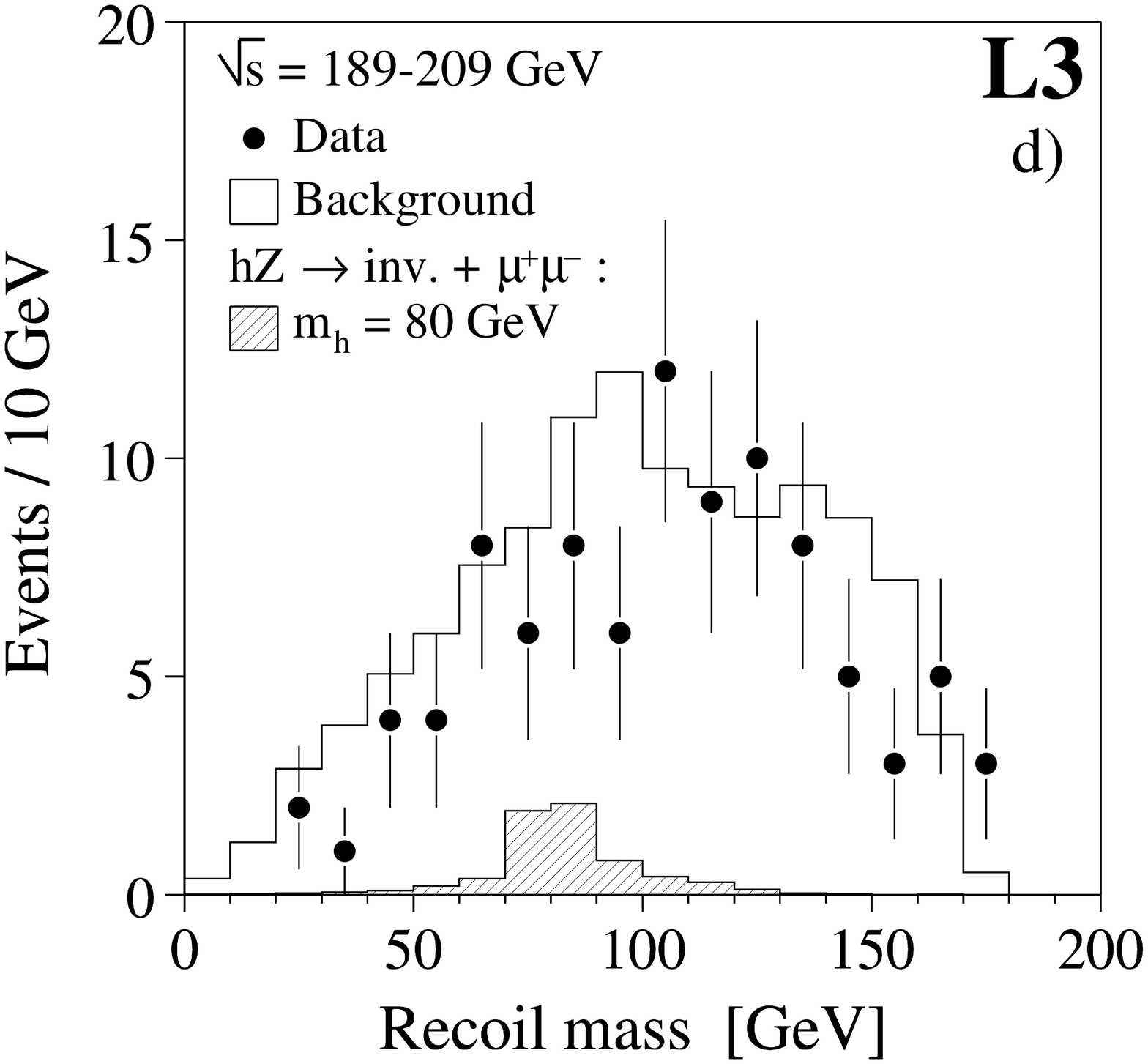,width=0.45\textwidth}}\\
      \mbox{\epsfig{figure=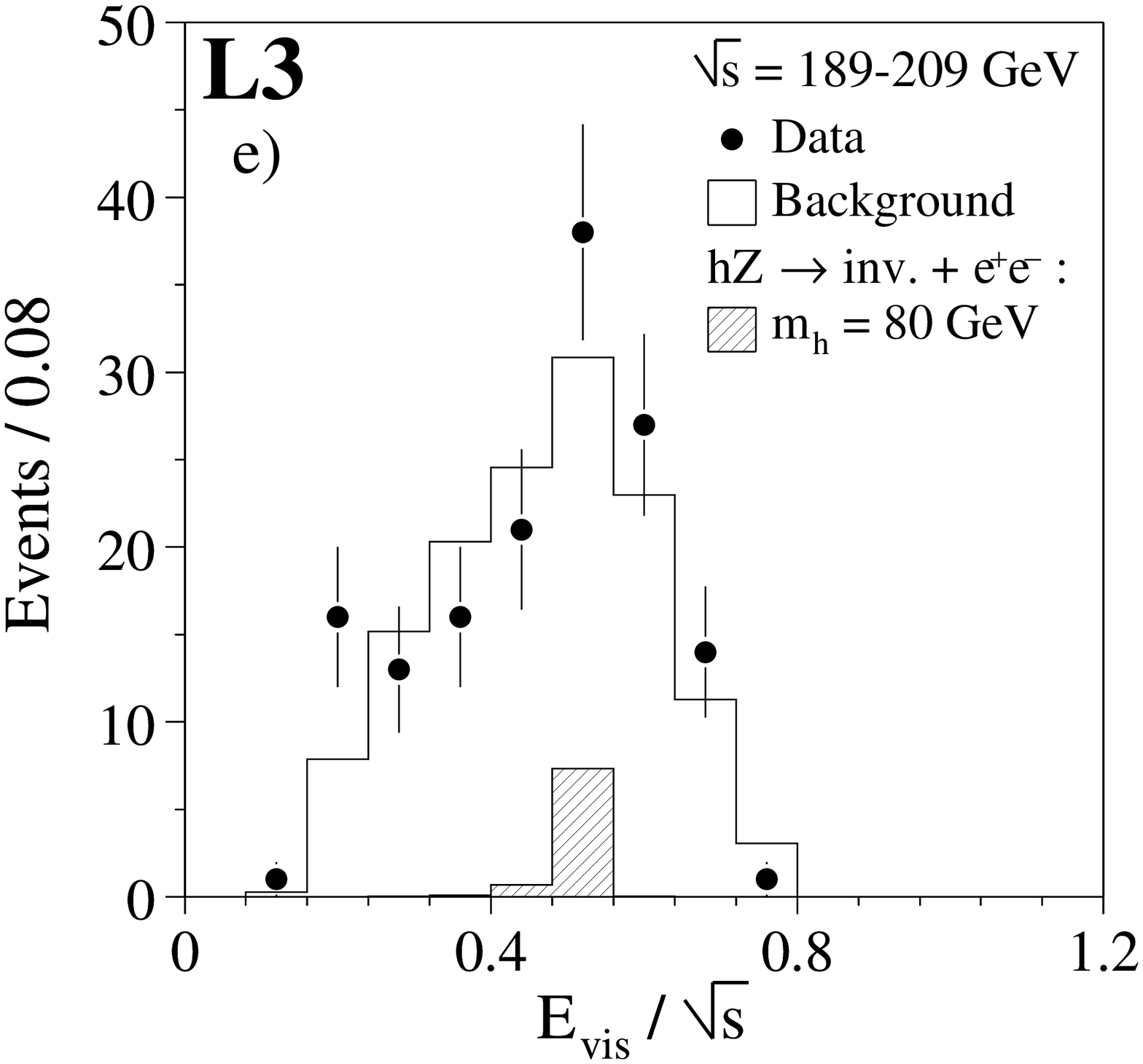,width=0.45\textwidth}}&
      \mbox{\epsfig{figure=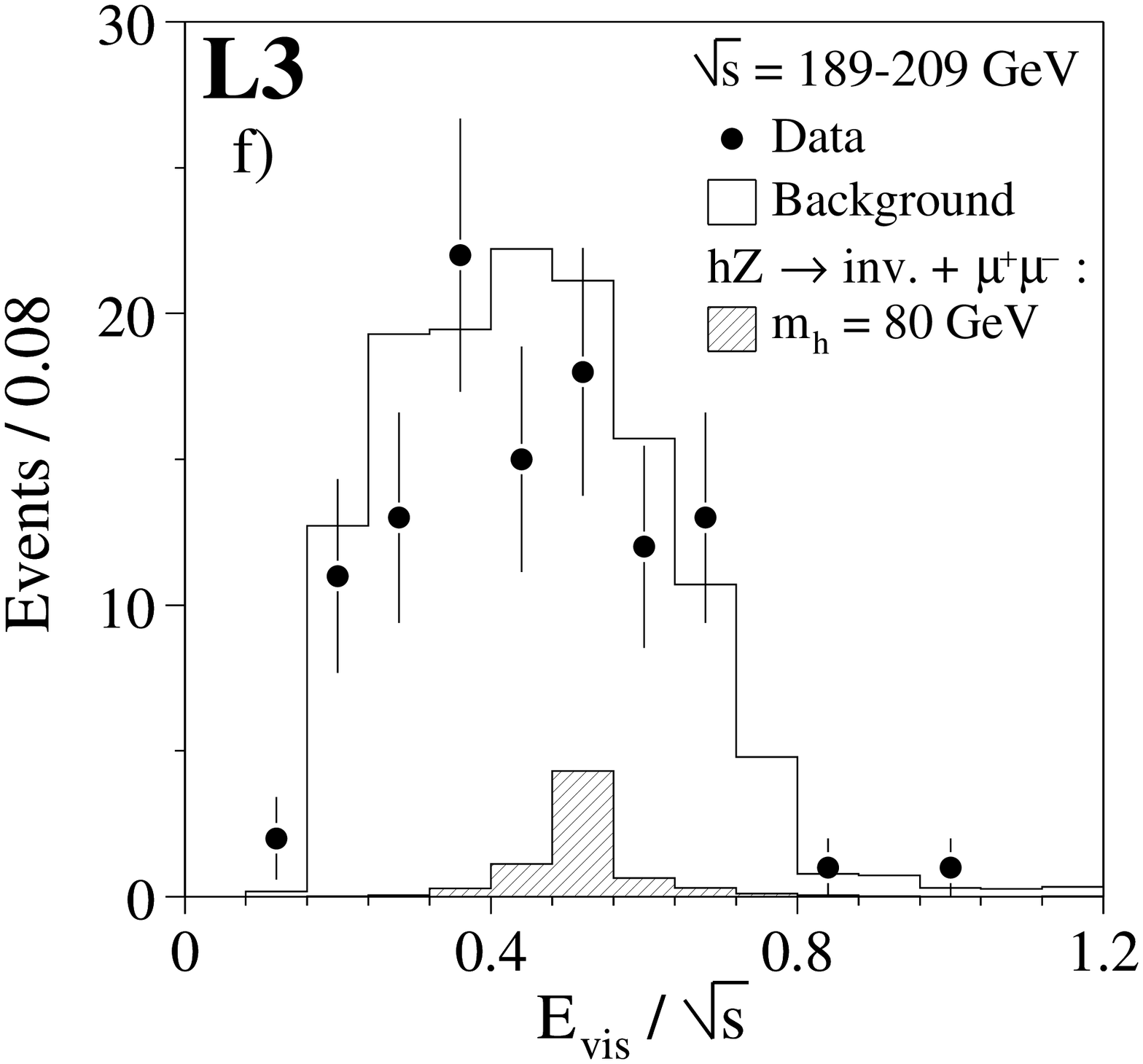,width=0.45\textwidth}}\\
    \end{tabular}
    \icaption[]{\label{fig:3}  Distributions of the visible and recoil
    masses and of the visible energy for events with
    electrons or muons and missing energy. The dots
    represent the data, the open area the sum of all background
    contributions and the hatched histogram the expectation for a
    signal. The selection criteria on the visible masses are illustrated by
    the arrows.}
  \end{center}
\end{figure}

A total of 147 electron pairs and 115 muon pairs are selected, in good
agreement with the Standard Model expectation of 136 and 130 events,
respectively. These events are mostly due to four-fermion production,
as summarised in Table~\ref{tab:emu}. Signal efficiencies depend on
$m_{\rm h}$ and are about 60\% and 50\% for final states with
electrons and muons, respectively.  The distributions after this
preselection of the visible and recoil masses of the lepton pairs, as
well as of the visible energy of the events are shown in
Figures~\ref{fig:3}. A good agreement between data and Monte Carlo
expectations is found.

The main criteria to isolate signal events is to require the
consistency of the visible mass with the mass of the Z boson. Two
ranges are chosen, $86\GeV-95\GeV$ for electrons and $80\GeV-99\GeV$
for muons, as illustrated in Figures~\ref{fig:3}a and~\ref{fig:3}b.
In addition, the event selection requires $|\cos{\theta_{miss}}| <
0.9$.  In order to reduce the four-fermion background and increase the
signal sensitivity, events are classified according to the value of
the recoil mass. If it is below $85\GeV$, a light-Higgs selection is
further applied. A heavy-Higgs selection is applied otherwise. The
light-Higgs selection relies on three cuts, common to both final
states: $\zeta_\ell >100^\circ$, $E_{vis}/\sqrt{s}<0.57$ and
$p_{z}/\sqrt{s}<0.25$, where $p_{z}$ is the projection of the event
momentum along the direction of the beams. In addition, events with
muons are required to satisfy $p_{t}>14\GeV$. The heavy-Higgs
selection requires $E_{vis}/\sqrt{s}<0.45$ for both final states and
$p_{t}>20\GeV$ for final states with muons.

After these cuts, a total of 6 events are observed in the electron
final-state and 9 in the muon final state, consistent with the
Standard Model background expectations of 9.7 and 11.1 events,
respectively, largely due to four-fermion final states. These results
are summarised in Table~\ref{tab:emu} while the signal efficiencies
are detailed in Table~\ref{tab:effi}. The distributions of the visible
mass of the events, after all other cuts are applied, are shown in
Figures~\ref{fig:4}a and~\ref{fig:4}b, while Figures~\ref{fig:4}c
and~\ref{fig:4}d show the distributions of the recoil mass. No
indication for a Higgs signal is found in these distributions.

%
\section{Results}
%

\begin{figure}[p]
  \begin{center}
    \begin{tabular}{cc}
      \mbox{\epsfig{figure=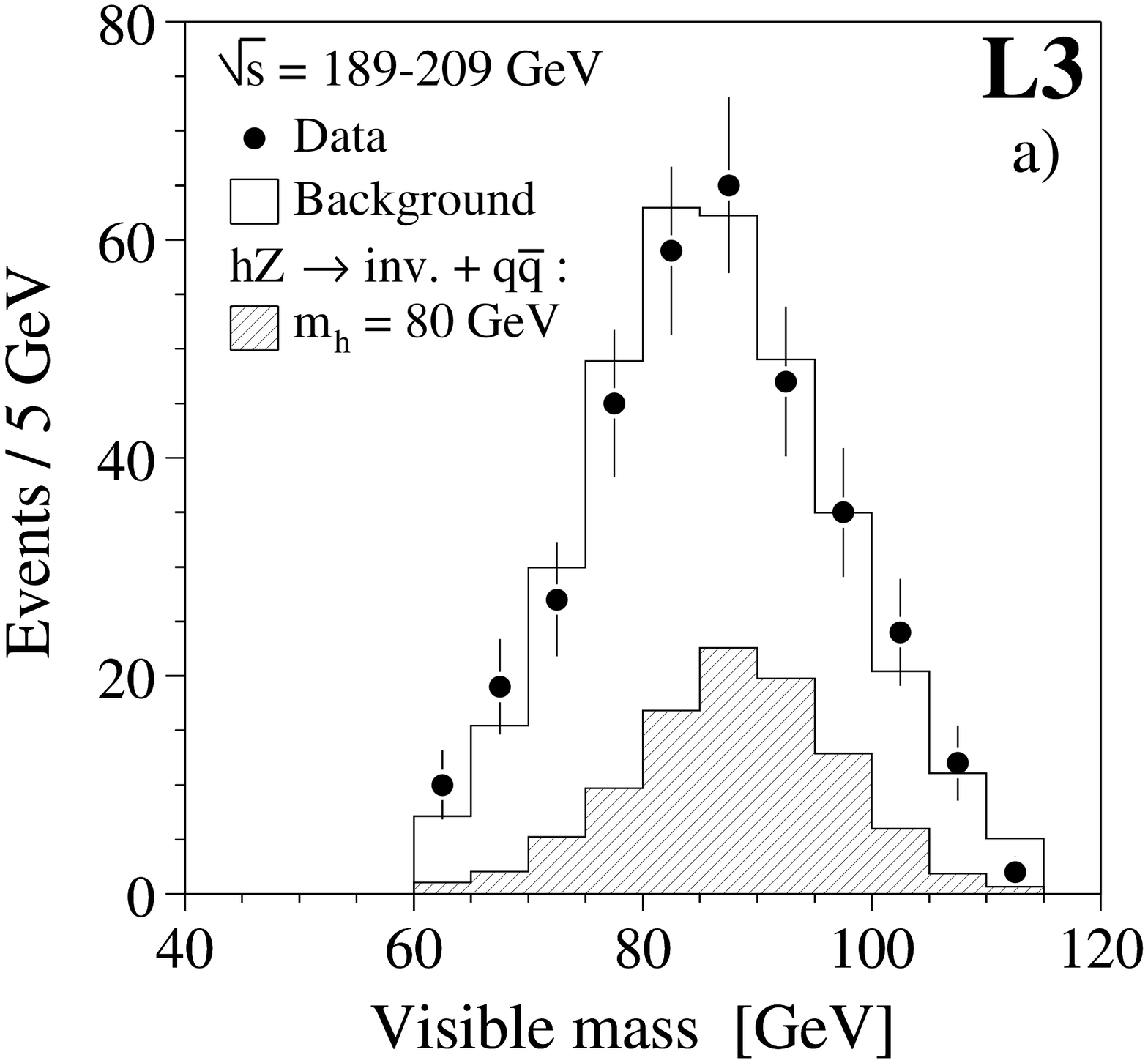,width=0.45\textwidth}}&
      \mbox{\epsfig{figure=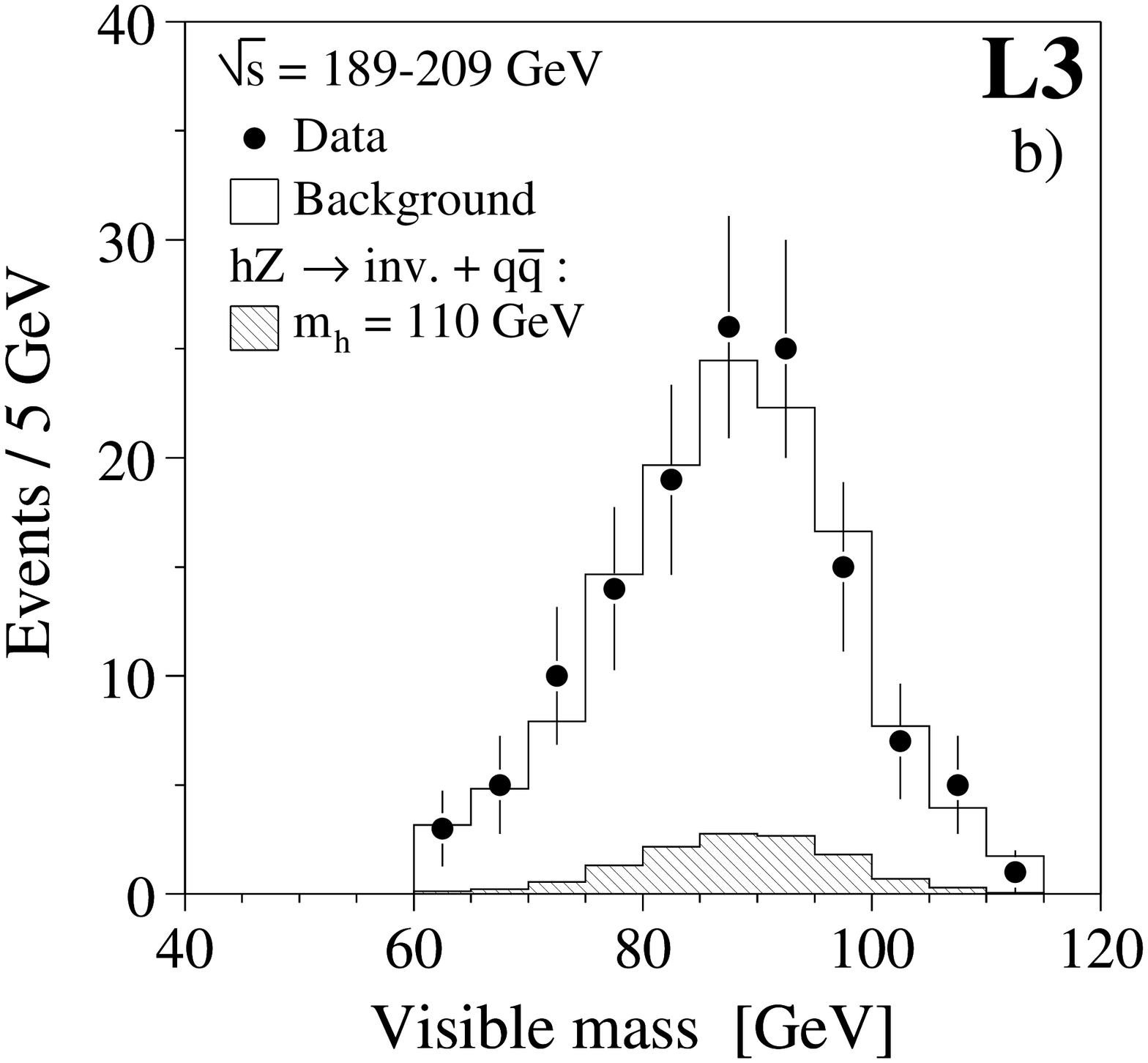,width=0.45\textwidth}}\\
      \mbox{\epsfig{figure=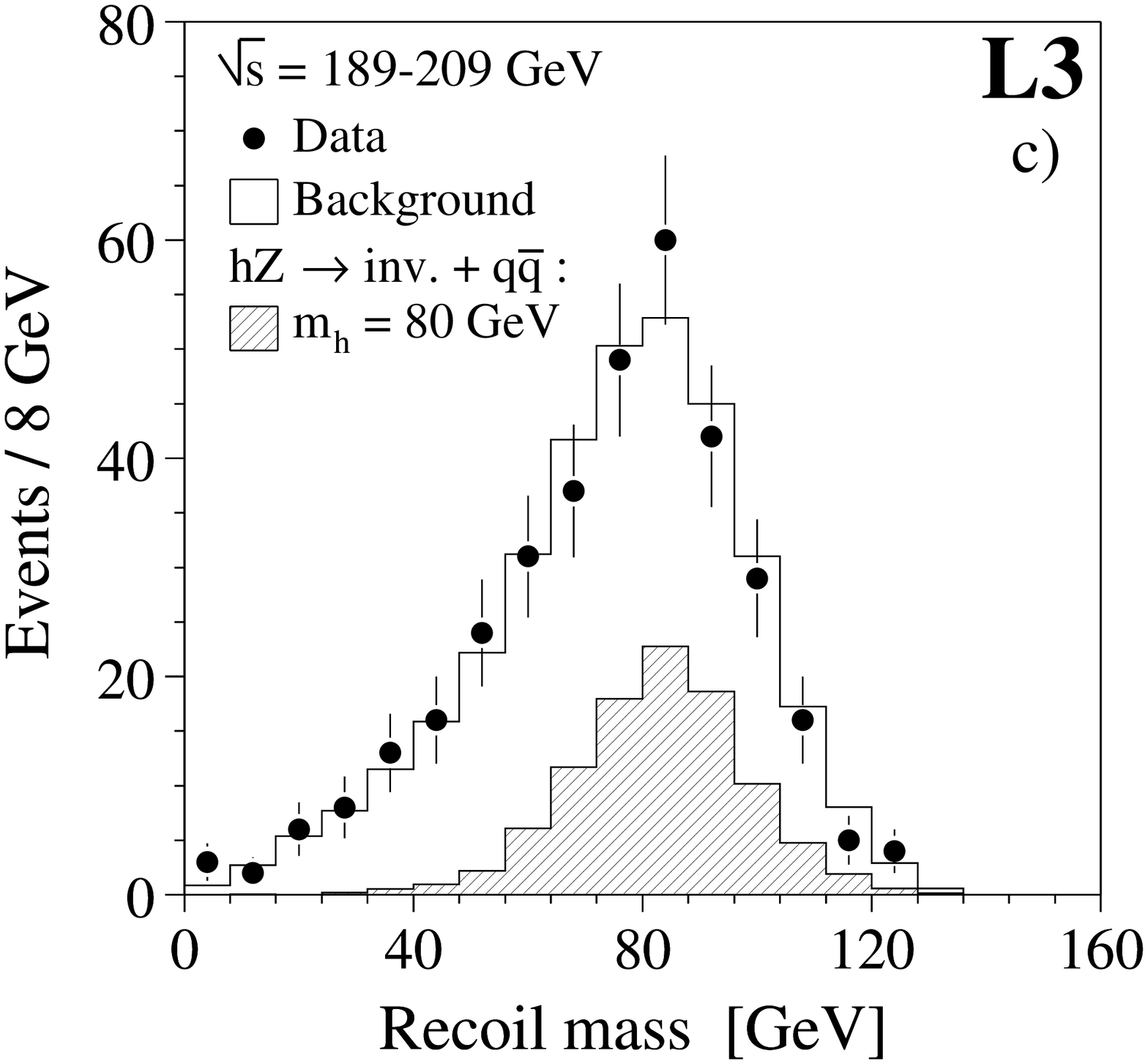,width=0.45\textwidth}}&
      \mbox{\epsfig{figure=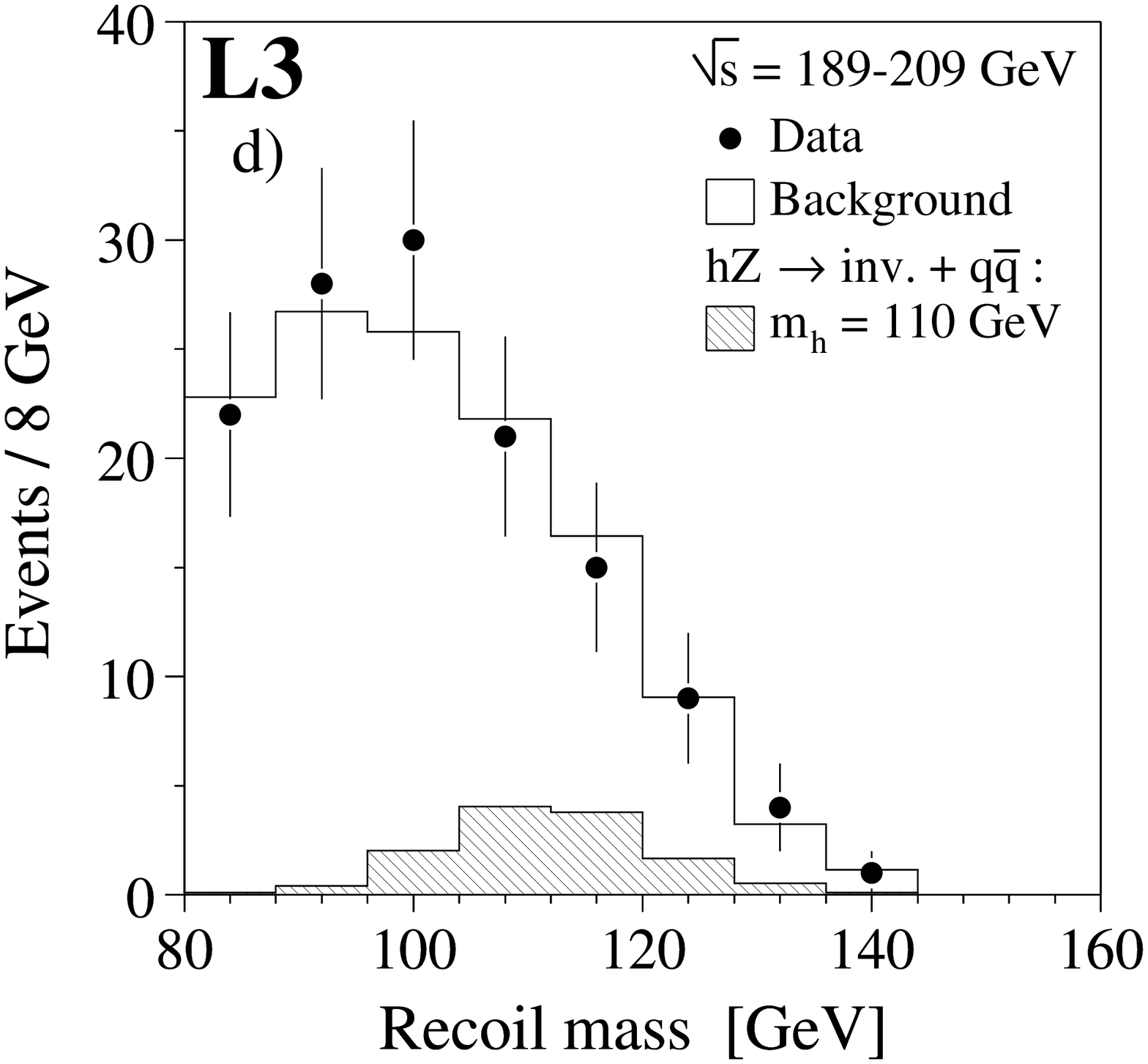,width=0.45\textwidth}}\\
      \mbox{\epsfig{figure=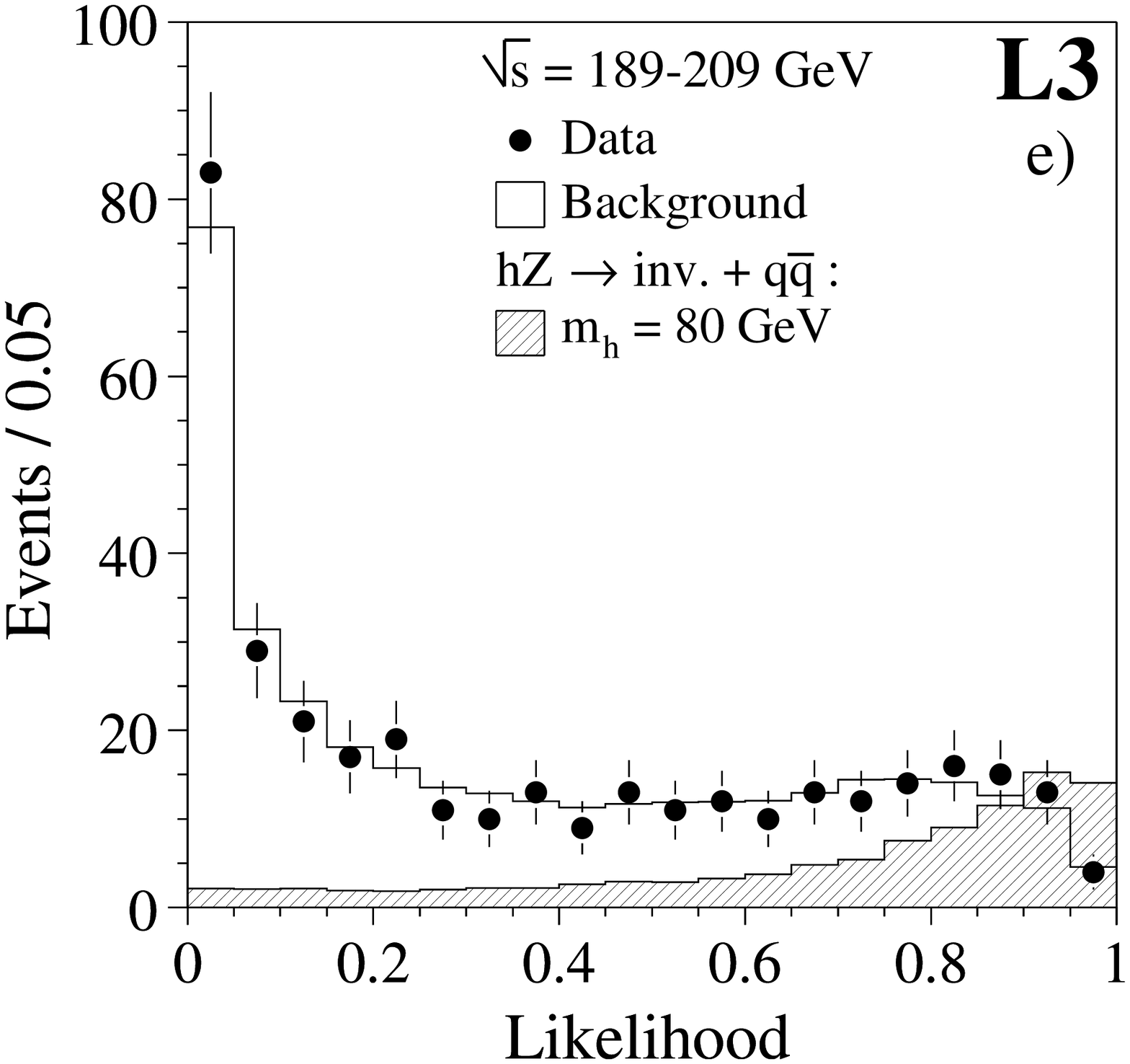,width=0.45\textwidth}}&
      \mbox{\epsfig{figure=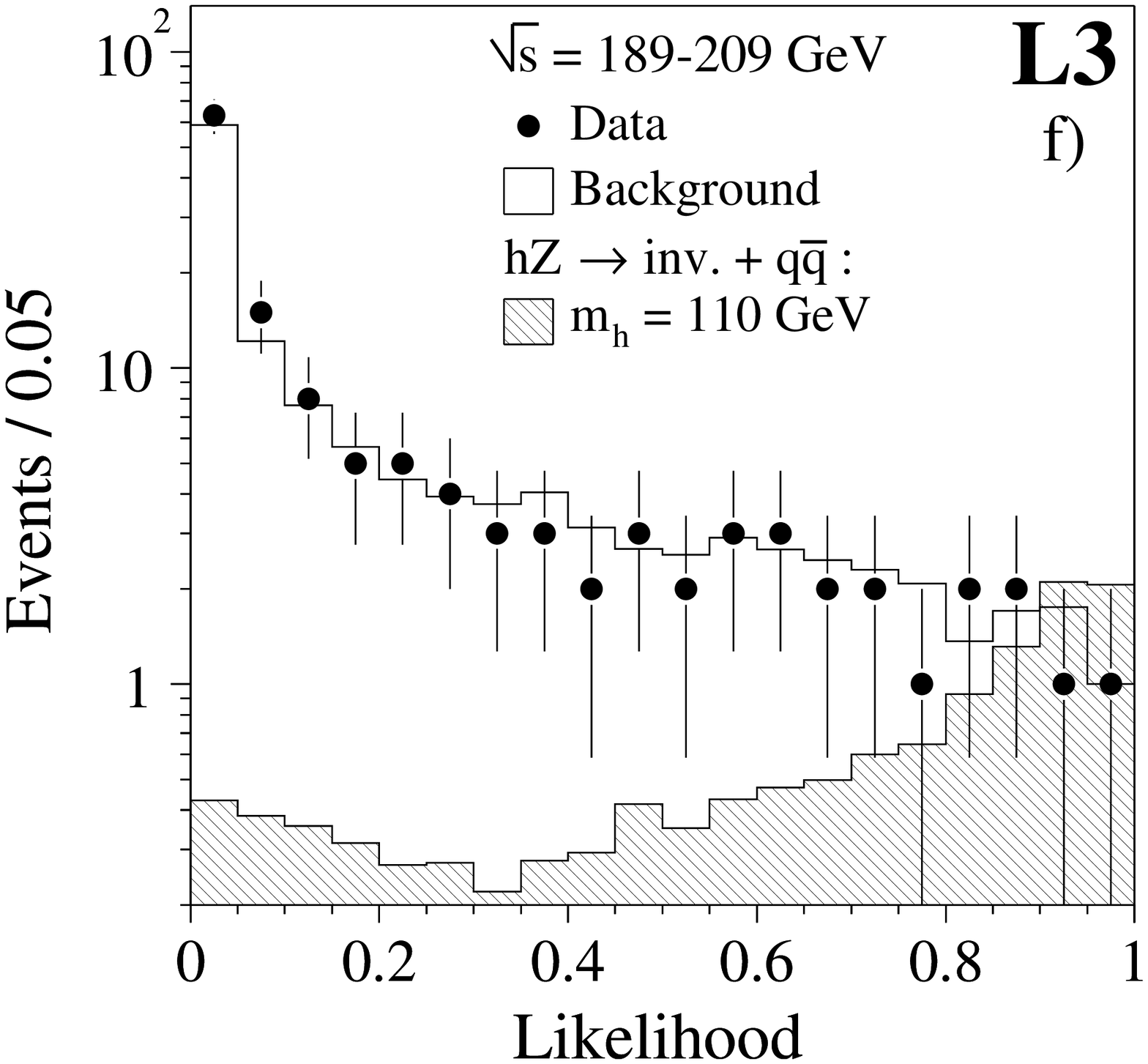,width=0.45\textwidth}}\\
    \end{tabular}
    \icaption[]{\label{fig:2} Distributions of the visible and recoil
    masses for events selected by the light- and heavy-Higgs
    analyses of events with hadrons and missing energy. The dots
    represent the data, the open area the sum of all background
    contributions and the hatched histogram the expectation for a
    signal. The distributions of the final discriminants used in the
    analysis are also shown.}
  \end{center}
\end{figure}

\begin{figure}[p]
  \begin{center}
    \begin{tabular}{cc}
      \mbox{\epsfig{figure=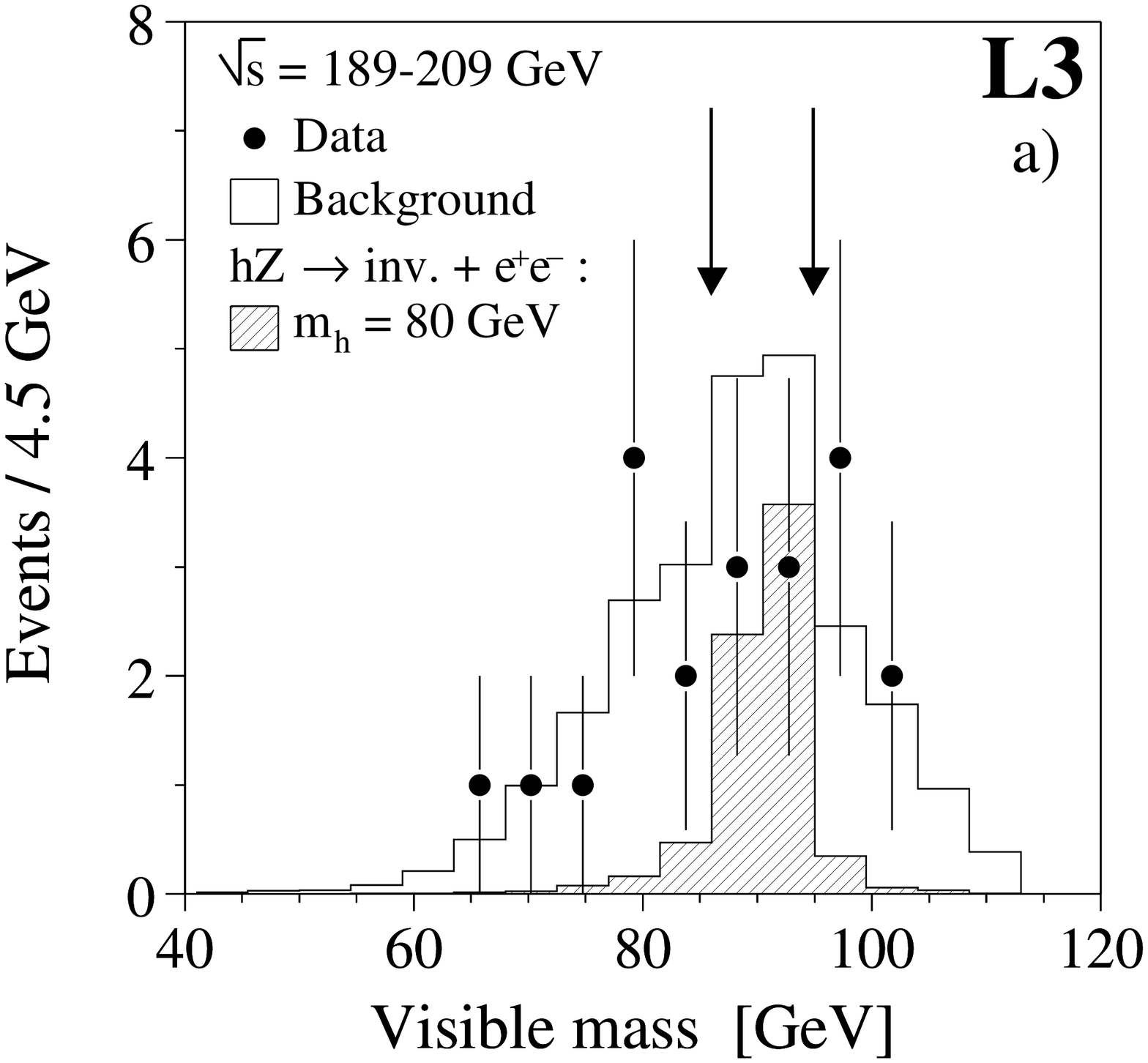,width=0.5\textwidth}}&
      \mbox{\epsfig{figure=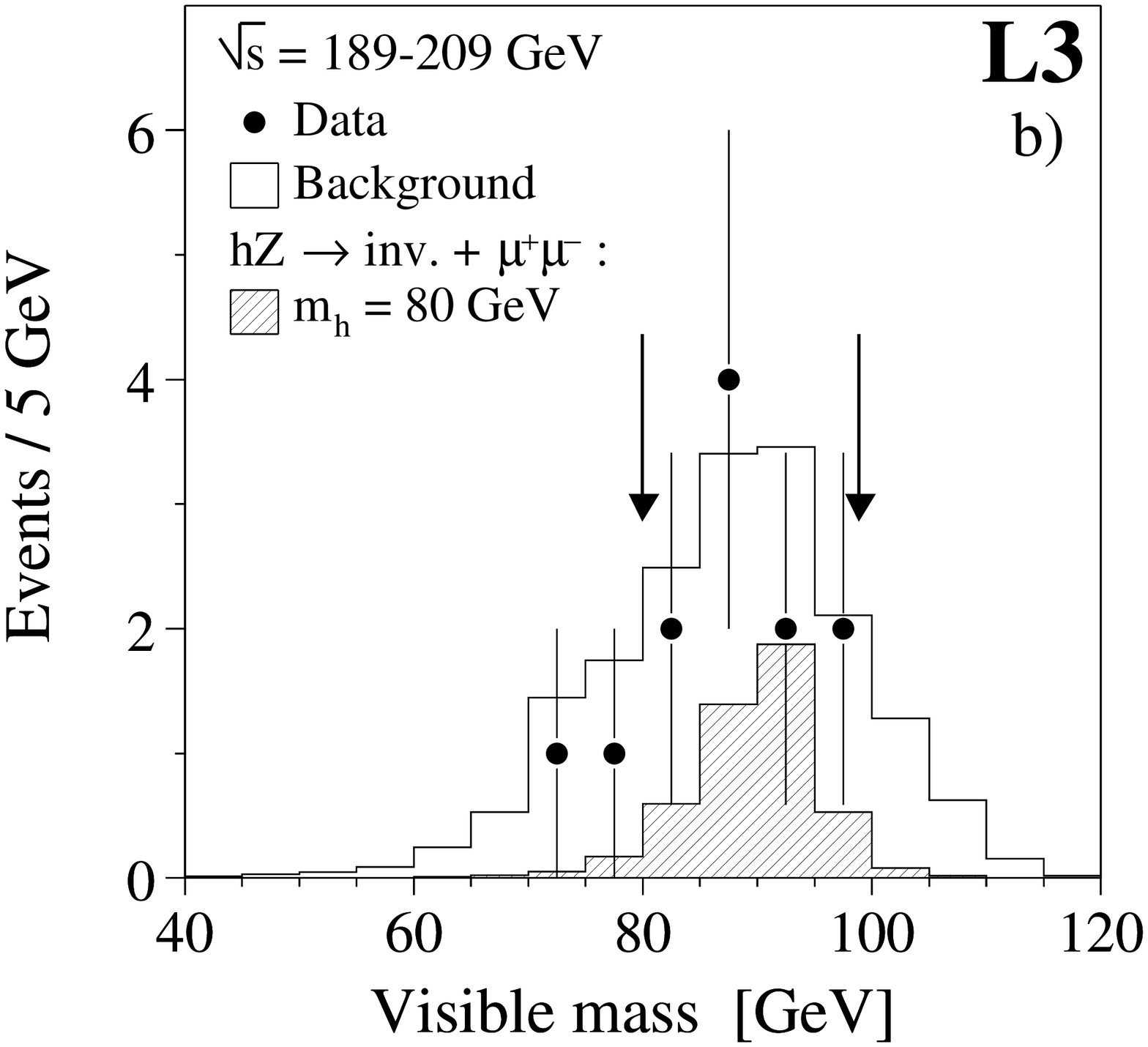,width=0.5\textwidth}}\\
      \mbox{\epsfig{figure=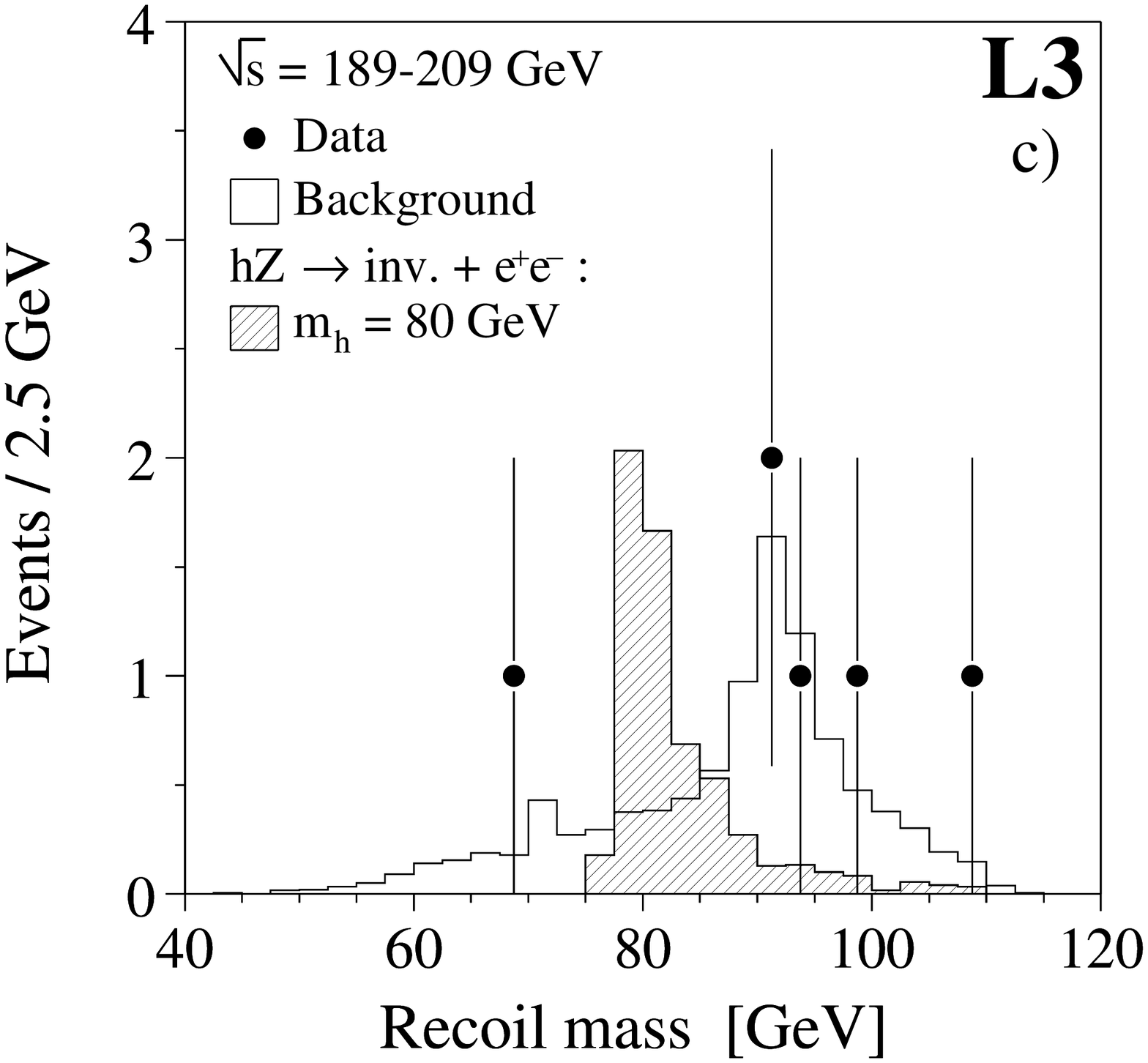,width=0.5\textwidth}}&
      \mbox{\epsfig{figure=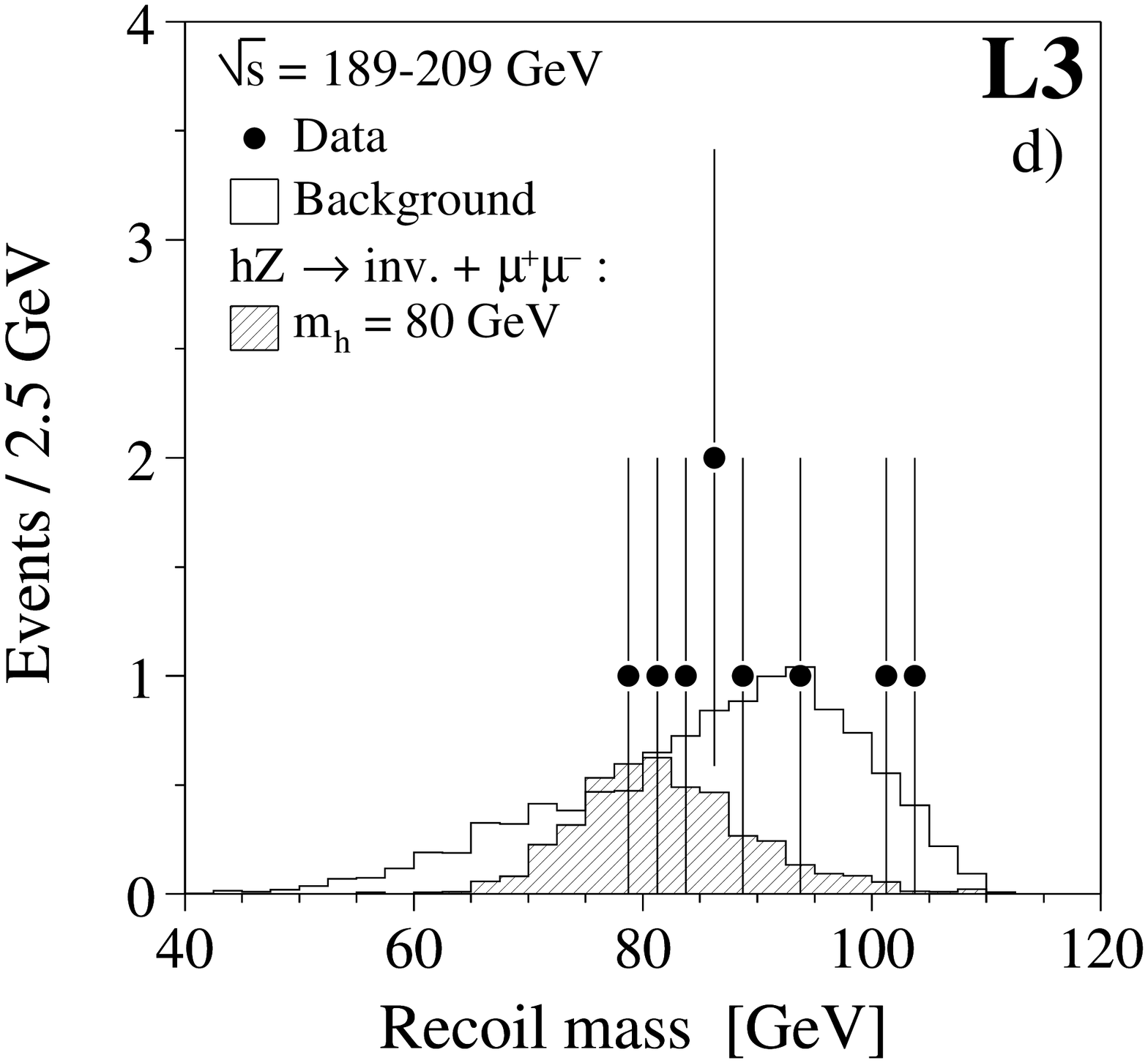,width=0.5\textwidth}}\\
    \end{tabular}
    \icaption[]{\label{fig:4} Distributions of the visible mass of
    events selected by the analysis of final states with a) electrons
    and missing energy and b) muons and missing energy after. The
    selection criteria on the visible masses are illustrated by the
    arrows. Distributions of the recoil mass after the application of
    all cuts are shown in c) for electrons and d) for muons. The dots
    represent the data, the open area the sum of all background
    contributions and the hatched histogram the expectation for a
    signal. }
  \end{center}
\end{figure}

\begin{figure}[p]
  \begin{center}
    \begin{tabular}{cc}
      \mbox{\epsfig{figure=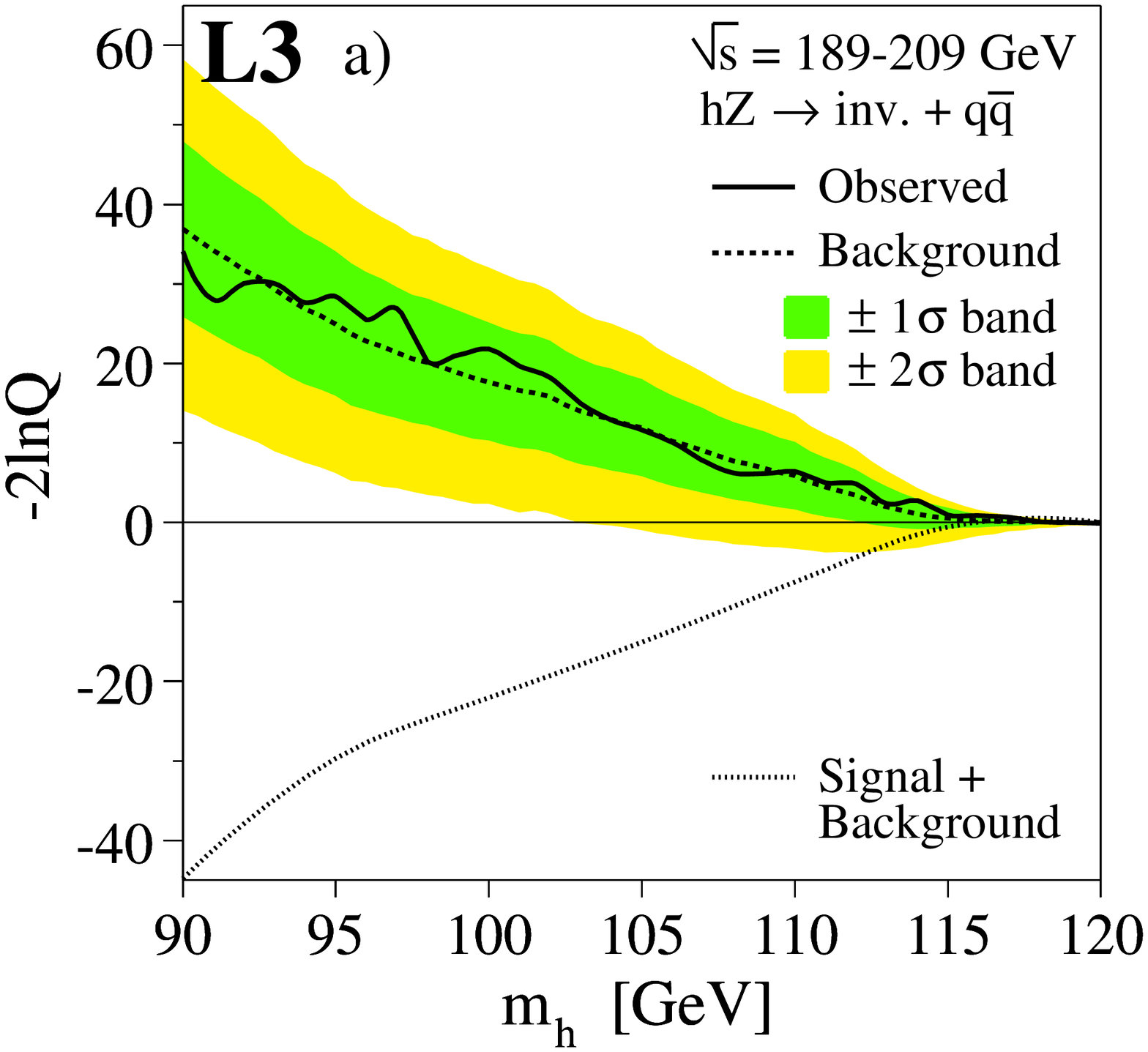,width=0.45\textwidth}}&
      \mbox{\epsfig{figure=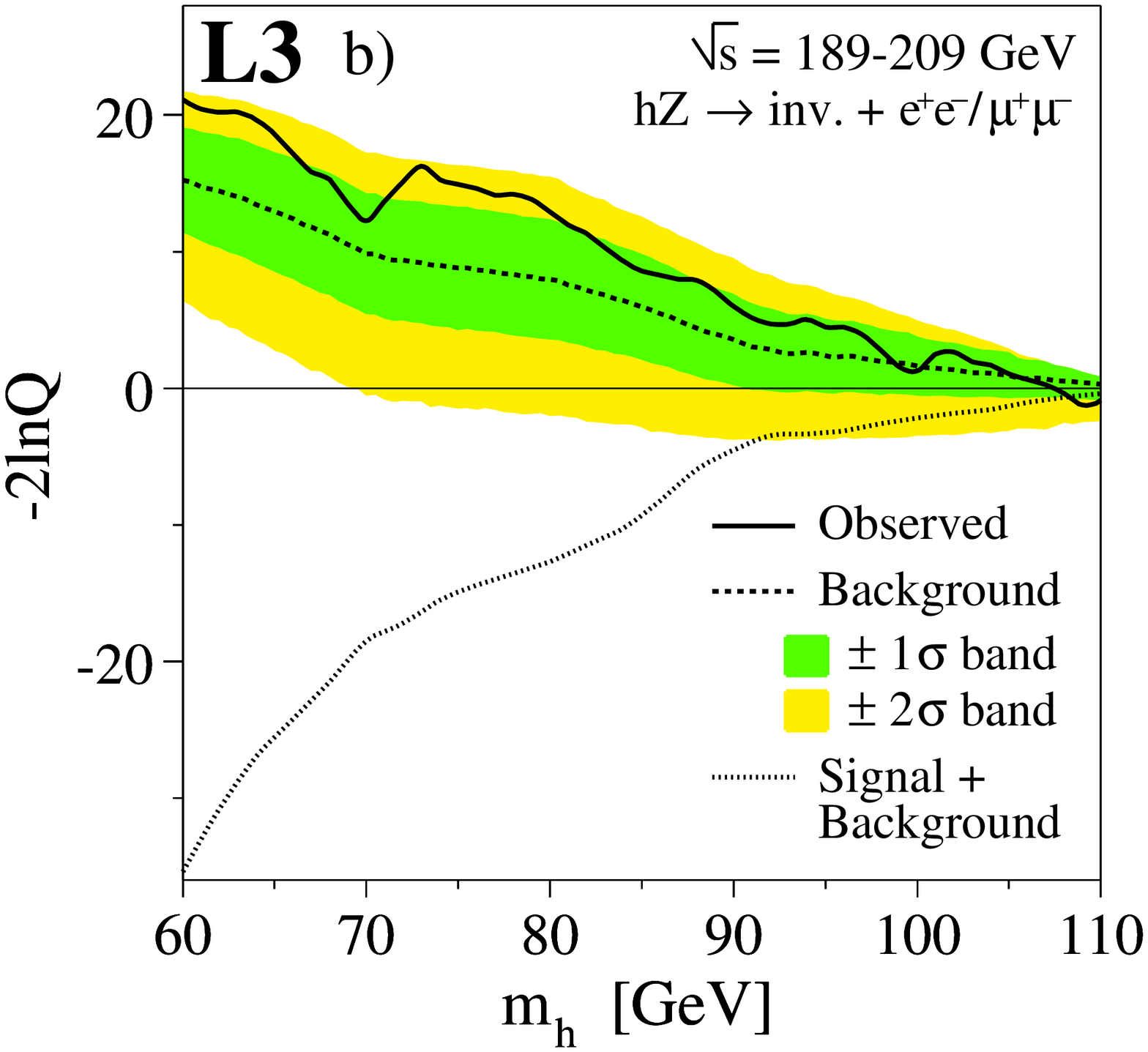,width=0.45\textwidth}}\\
      \mbox{\epsfig{figure=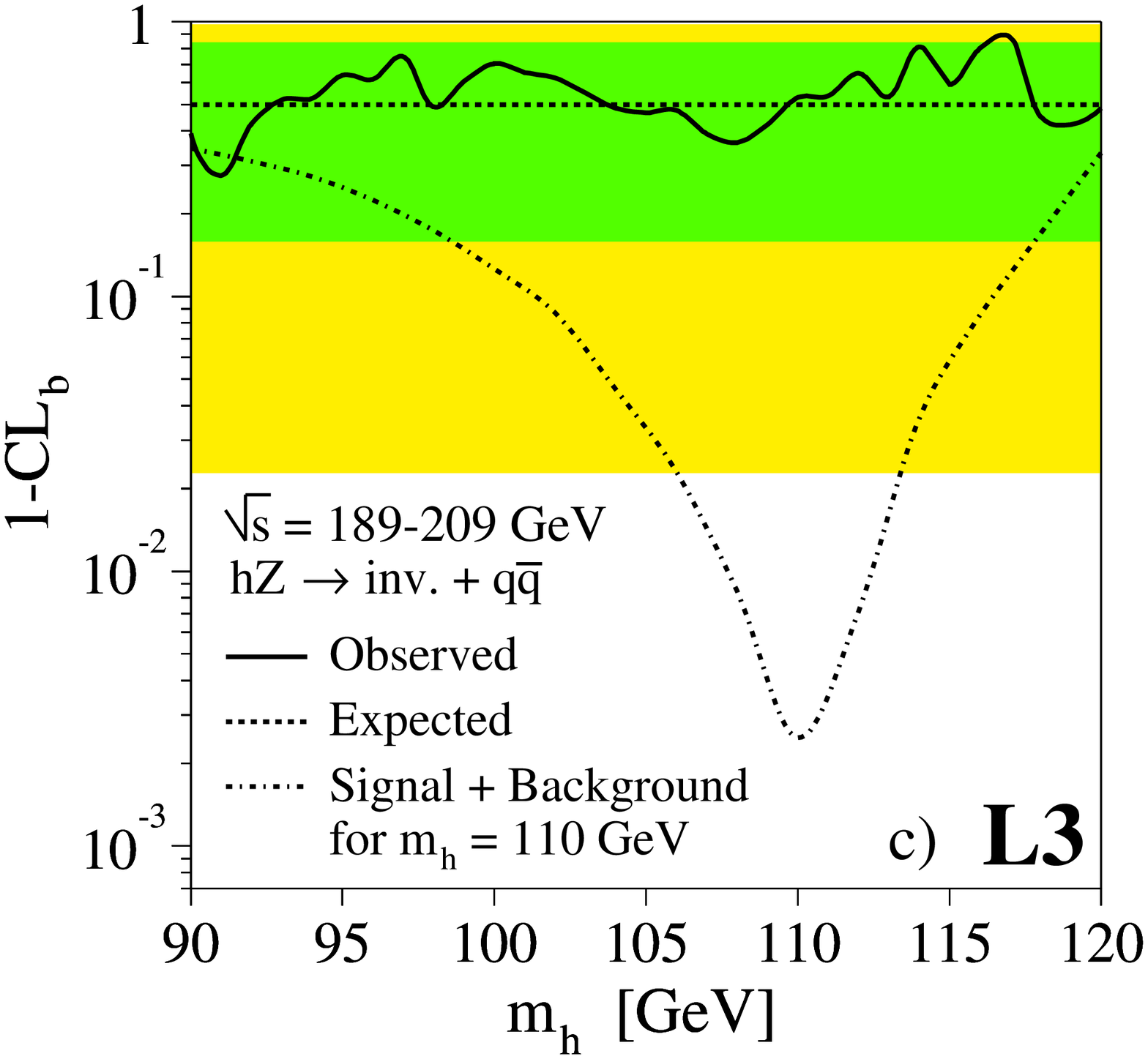,width=0.45\textwidth}}&
      \mbox{\epsfig{figure=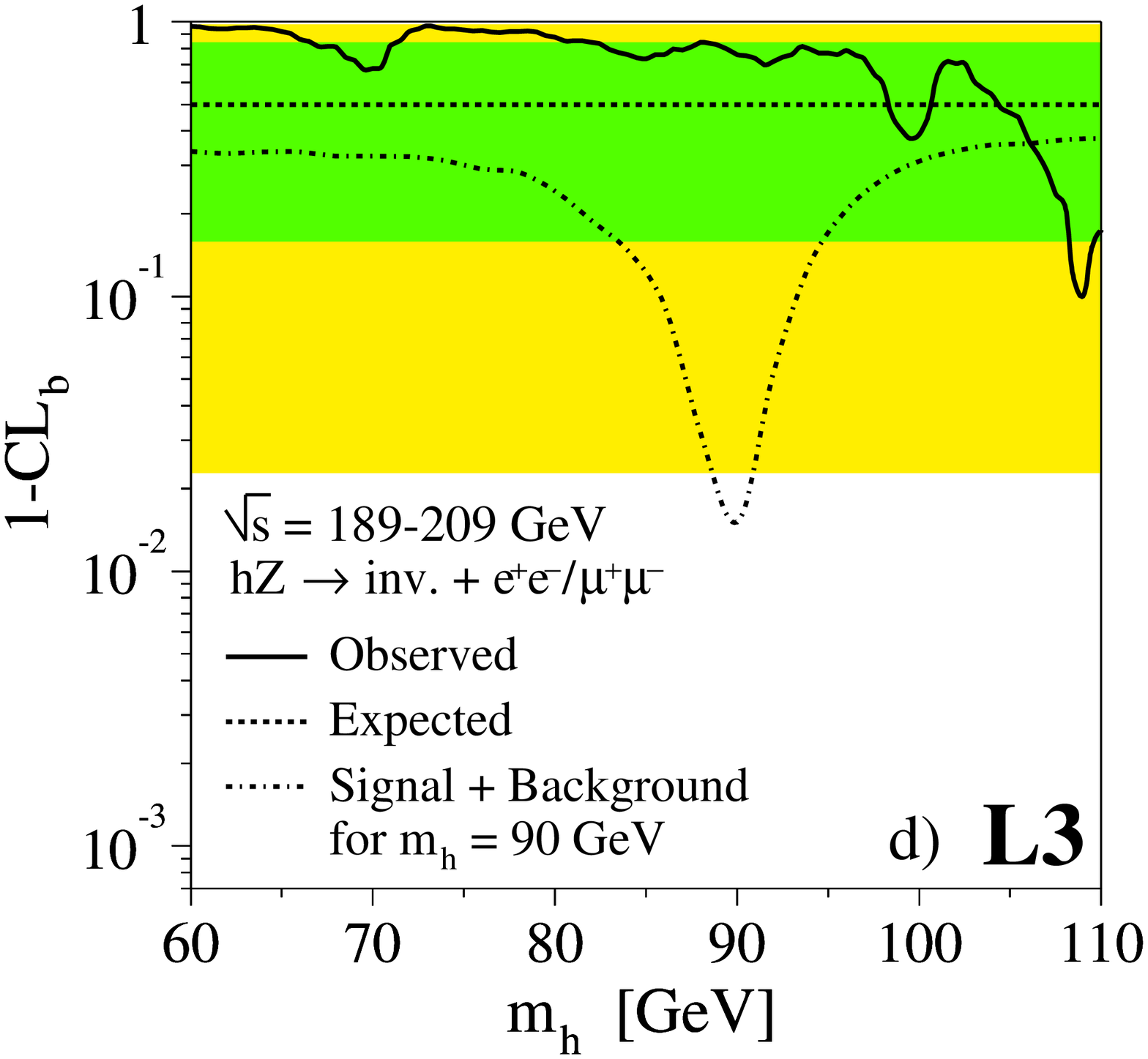,width=0.45\textwidth}}\\
      \mbox{\epsfig{figure=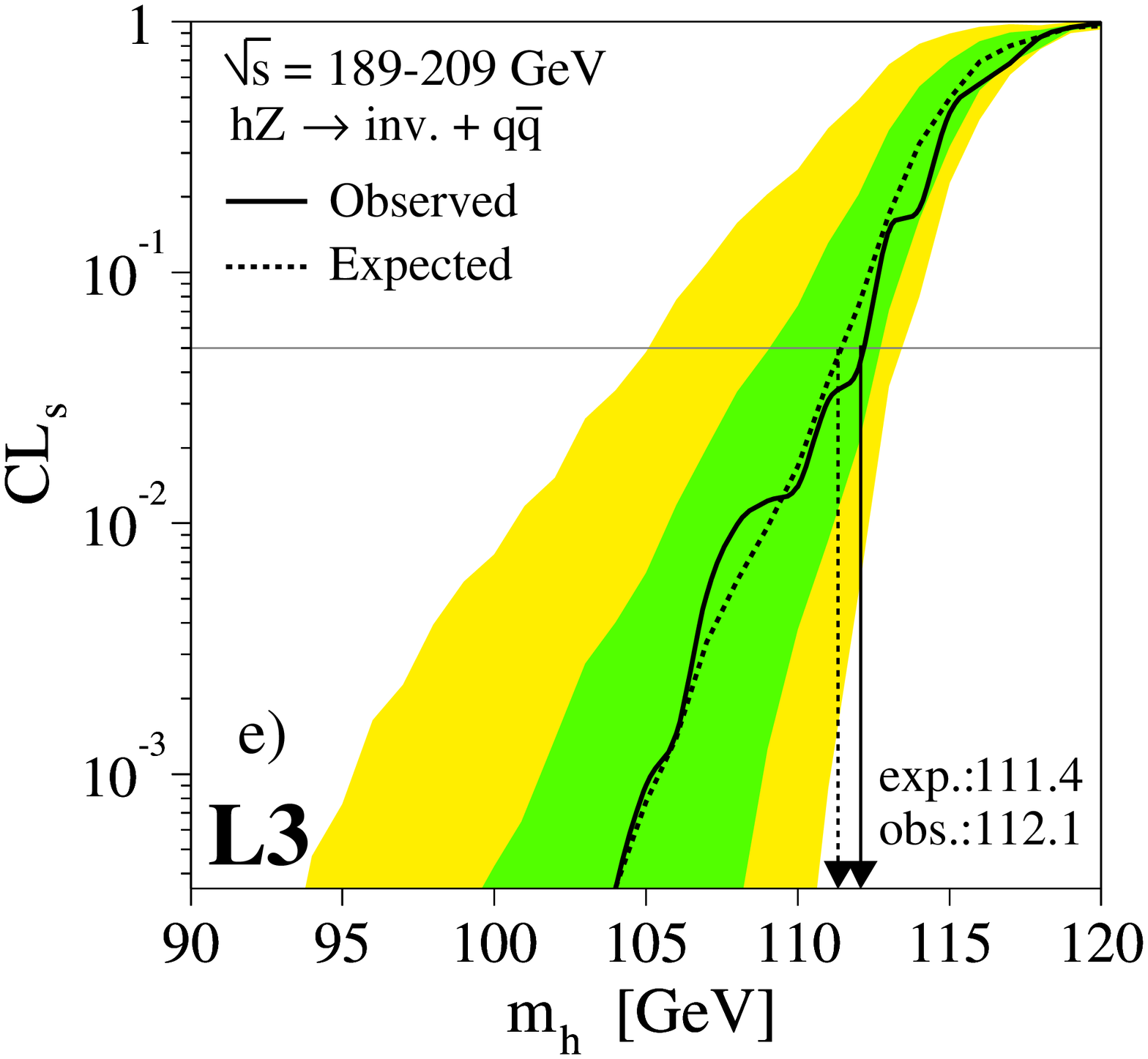,width=0.45\textwidth}}&
      \mbox{\epsfig{figure=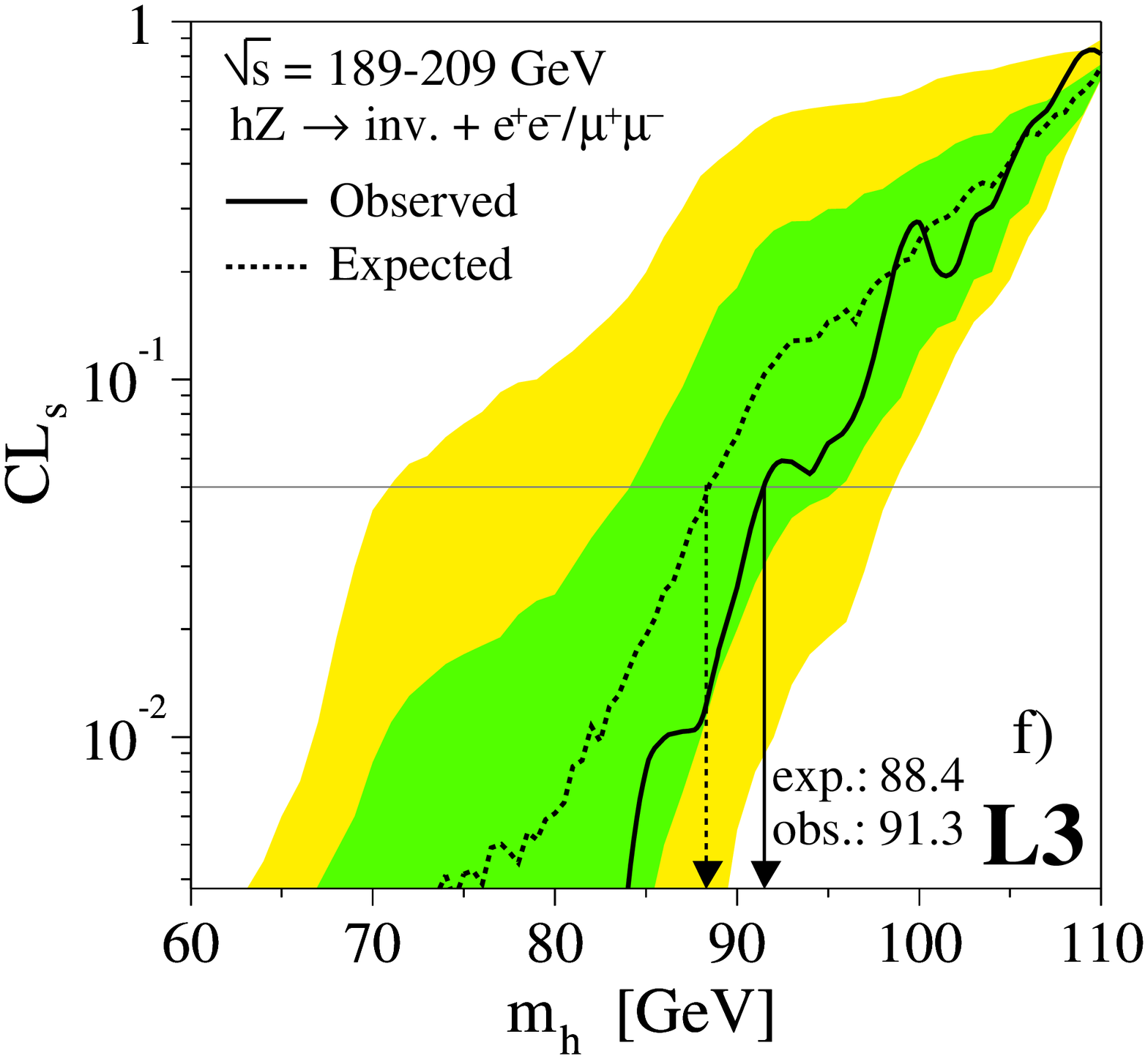,width=0.45\textwidth}}\\
    \end{tabular}
    \icaption[]{\label{fig:5}  Distributions as a function of $m_{\rm
      h}$ of the log-likelihood ratio for a) the hadron and b) the
      lepton analyses; of the $1-\rm{CL}_b$ estimator for c) the
      hadron and d) the lepton analyses; of the $\rm{CL}_s$ estimator
      for e) the hadron and f) the lepton analyses, together with the
      expected and observed lower limits on $m_{\rm h}$.}
  \end{center}
\end{figure}

No evidence is found for a signal due to the production of
invisibly-decaying Higgs bosons in association with a Z boson decaying
into hadrons, electrons or muons either in the total counts of events
or in the distributions of the discriminant variables and the recoil
masses. The results of this search are therefore expressed in terms of
limits on $m_{\rm h}$. In the hypothesis that an invisibly-decaying
Higgs boson is produced with the same cross section of the Standard
Model Higgs boson, a technique based on a log-likelihood
ratio~\cite{lnq} is used to calculate the confidence level
$1-\rm{CL}_b$ that the observed events are consistent with background
expectation. The distributions of the final discriminating variables
of the hadron selection, presented in Figures~\ref{fig:2}e
and~\ref{fig:2}f, and of the recoil masses to the lepton system,
presented in Figures~\ref{fig:4}c and~\ref{fig:4}d, are used in the
calculation which yields the results presented in Figure~\ref{fig:5}
for the hadron and lepton analyses in terms of the log-likelihood
ratio and $1-\rm{CL}_b$ as a function of $m_{\rm h}$. No structure
which could hint to the presence of a signal is observed. The
confidence level for the presence of the expected signal~\cite{lnq},
$\rm{CL}_s$, is also depicted in Figure~\ref{fig:5} for both analyses,
as a function of $m_{\rm h}$. Lower limits at 95\% confidence level
(CL) on $m_{\rm h}$ are derived from the results of the hadron and
lepton analyses as $112.1\GeV$ and $91.3\GeV$, respectively, in good
agreement with the expected limits of $111.4\GeV$ and $88.4\GeV$.

These limits include the systematic uncertainties on the signal
efficiency and the background normalisation listed in
Table~\ref{tab:syst}. These follow from the limited Monte Carlo
statistics and from the uncertainties on the cross sections of the
background processes. Additional sources of systematic uncertainties,
collectively indicated as ``detector response'' comprise uncertainties
in the determination of the energy scale of the detector and possible
discrepancies between data and Monte Carlo in the tails of the
variables used in the event selection. The inclusion of the systematic
uncertainties lowers the limits by about $200 \MeV$

\begin{figure}[p]
  \begin{center}
    \begin{tabular}{cc}
      \mbox{\epsfig{figure=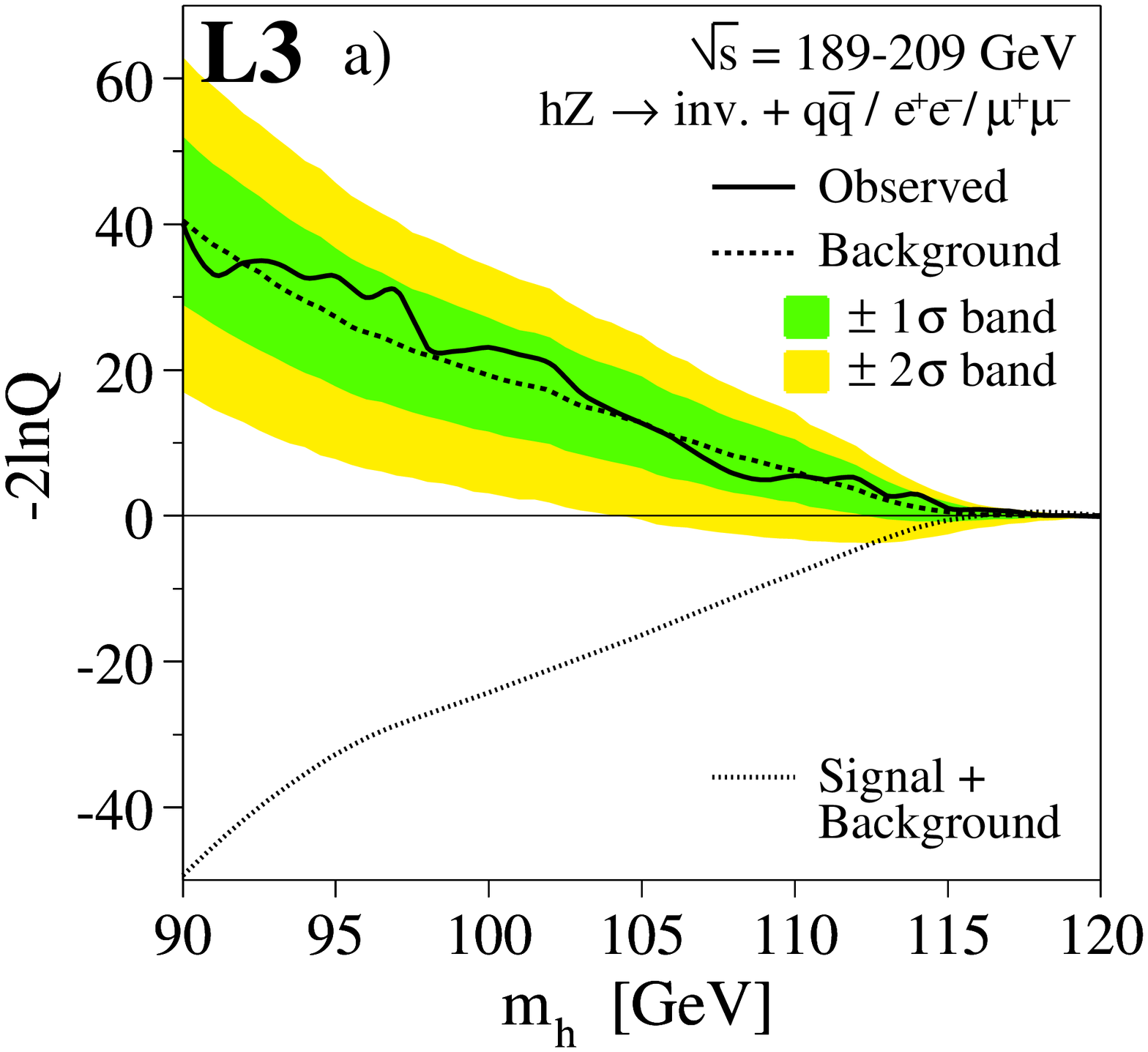,width=0.5\textwidth}}&
      \mbox{\epsfig{figure=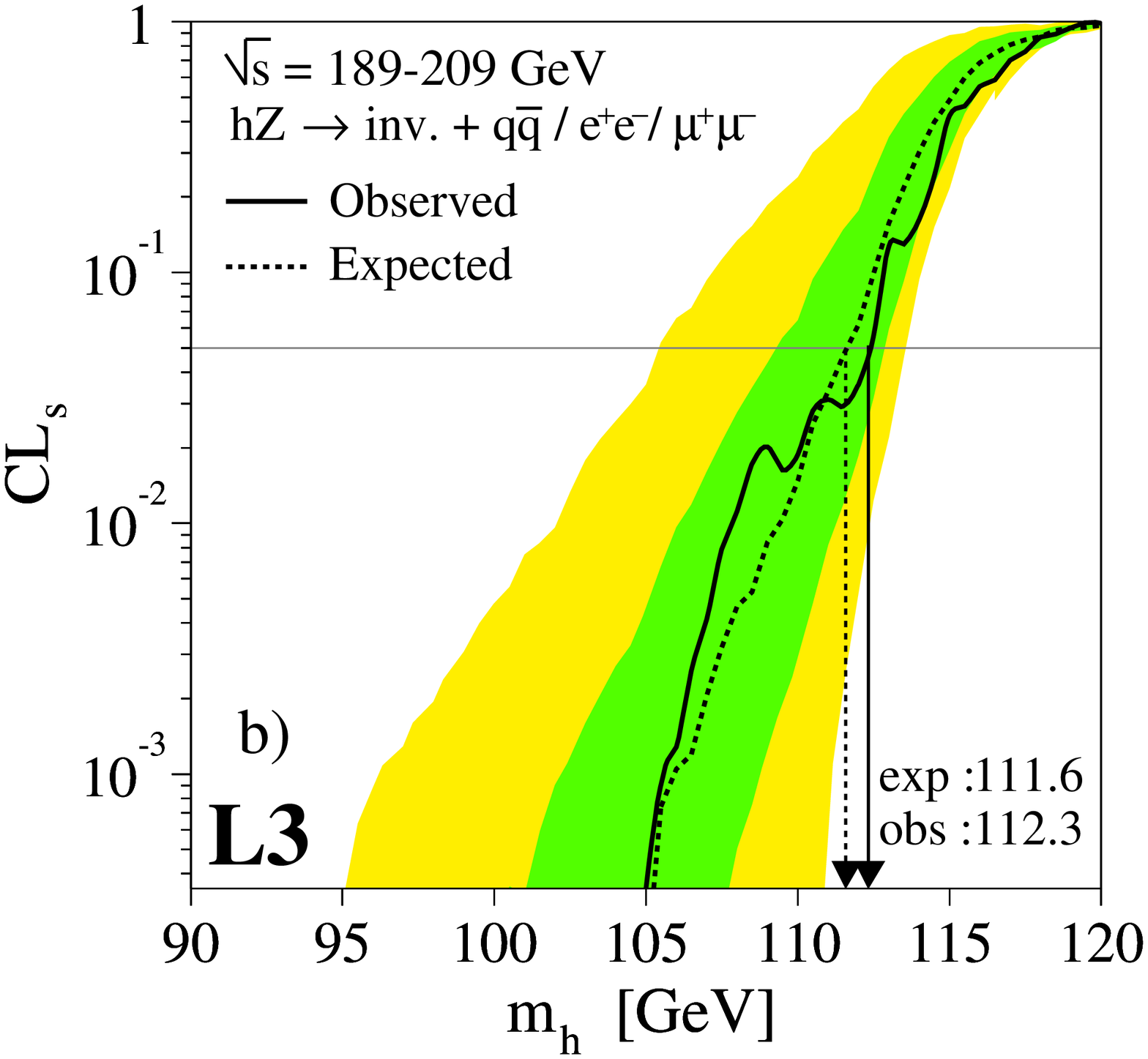,width=0.5\textwidth}}\\
      \mbox{\epsfig{figure=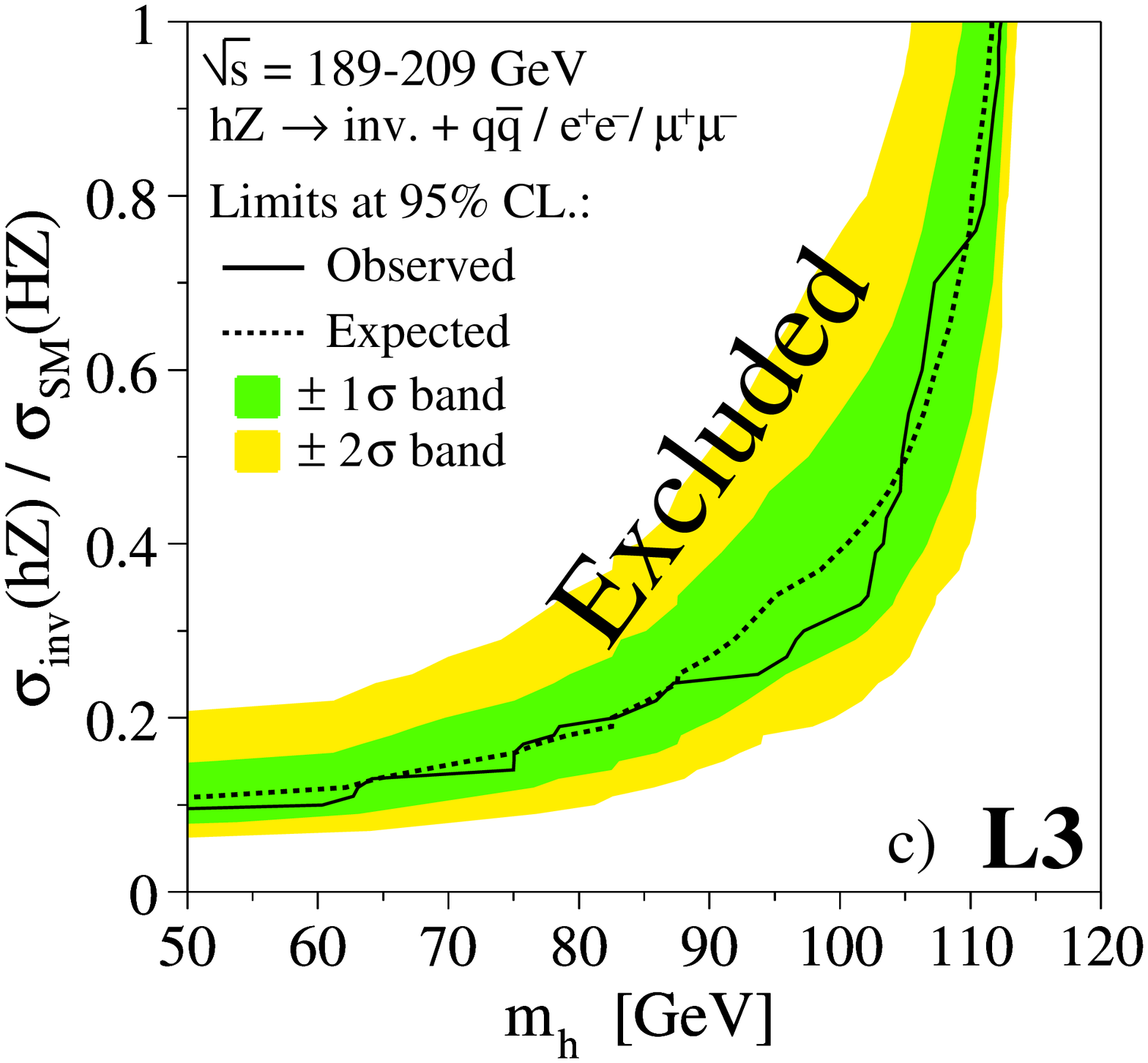,width=0.5\textwidth}}&
      \mbox{\epsfig{figure=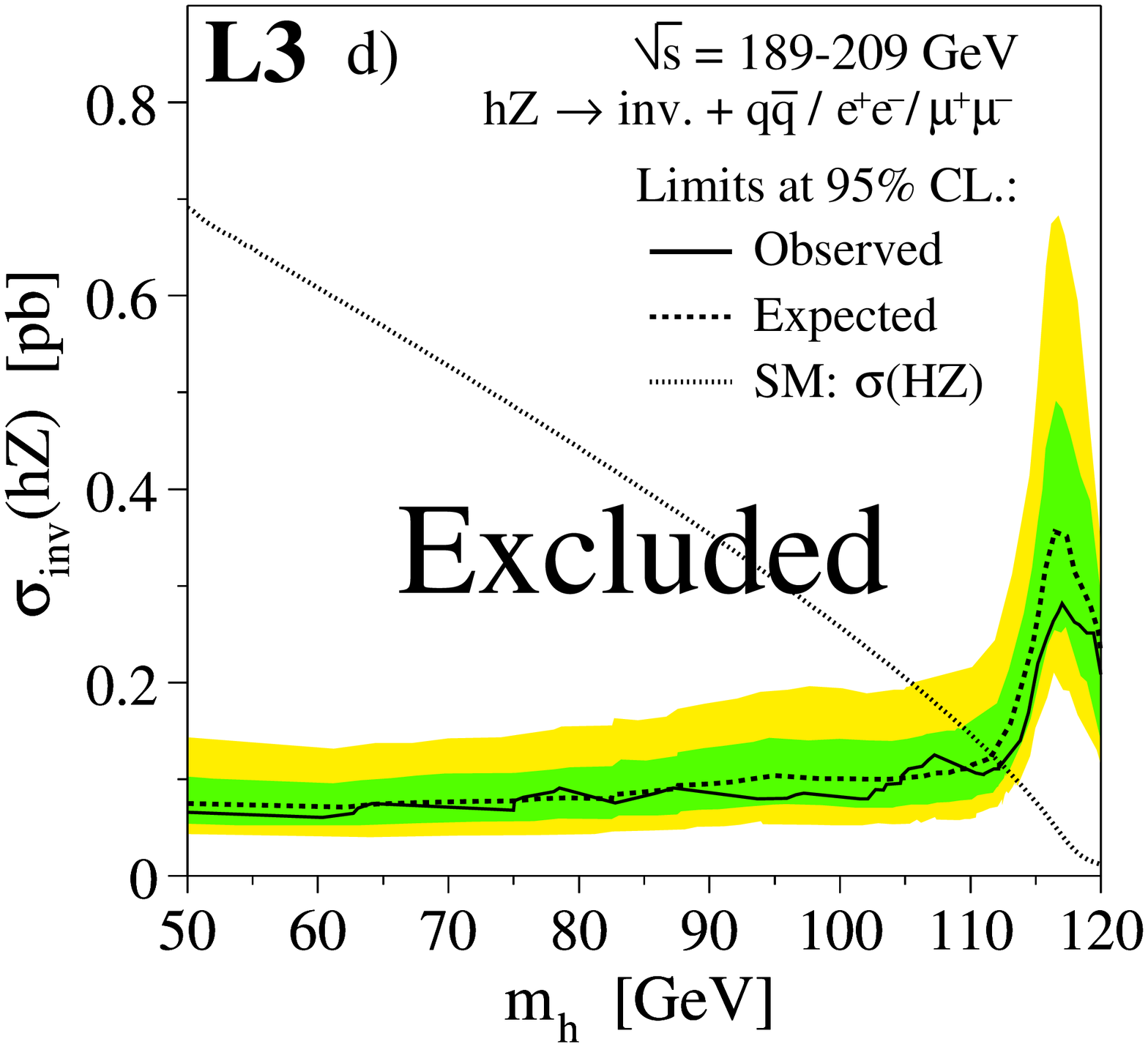,width=0.5\textwidth}}\\
    \end{tabular}
    \icaption[]{\label{fig:6} Distributions as a function of $m_{\rm
      h}$ for the combination of the hadron and lepton analyses of a)
      the log-likelihood; b) the $\rm{CL}_s$ estimator with the
      expected and observed lower limits on $m_{\rm h}$; c) upper
      limits on the ratio of the invisibly-decaying Higgs-boson cross
      section to the Standard Model one; d) upper limits on the cross
      section for the production of an invisibly-decaying Higgs
      boson.}
  \end{center}
\end{figure}

The results of the combination of the hadron and lepton selections is
expressed in terms of the log-likelihood ratio and $\rm{CL}_s$ as a
function of $m_{\rm h}$ shown in Figures~\ref{fig:6}a
and~\ref{fig:6}b.  A lower limit to the mass of an invisibly-decaying
Higgs boson is derived at 95\% CL as:
\begin{displaymath}
m_{\rm h} > 112.3\GeV,
\end{displaymath}
in good agreement with the expected limit of $111.6\GeV$. This limit
holds in the hypothesis that the invisibly-decaying Higgs boson is
produced with the same cross section of the Standard Model Higgs
boson. If this hypothesis is relaxed, upper limits as a function of
$m_{\rm h}$ are extracted on the ratio of the invisibly-decaying
Higgs-boson cross section to the Standard Model one, as shown in
Figure~\ref{fig:6}c. These limits are translated into the upper limits
on the cross section for the production of an invisibly-decaying Higgs
boson as a function of $m_{\rm h}$ shown in Figure~\ref{fig:6}d.

%

\newpage

%
%

\newpage
\typeout{   }     
\typeout{Using author list for paper 287 -  }
\typeout{$Modified: Jul 15 2001 by smele $}
\typeout{!!!!  This should only be used with document option a4p!!!!}
\typeout{   }
%
%
%
%
%
%

\newcount\tutecount  \tutecount=0
\def\tutenum#1{\global\advance\tutecount by 1 \xdef#1{\the\tutecount}}
\def\tute#1{$^{#1}$}
\tutenum\aachen            
\tutenum\nikhef            
\tutenum\mich              
\tutenum\lapp              
\tutenum\basel             
\tutenum\lsu               
\tutenum\beijing           
\tutenum\bologna           
\tutenum\tata              
\tutenum\ne                
\tutenum\bucharest         
\tutenum\budapest          
\tutenum\mit               
\tutenum\panjab            
\tutenum\debrecen          
\tutenum\dublin            
\tutenum\florence          
\tutenum\cern              
\tutenum\wl                
\tutenum\geneva            
\tutenum\hamburg           
\tutenum\hefei             
\tutenum\lausanne          
\tutenum\lyon              
\tutenum\madrid            
\tutenum\florida           
\tutenum\milan             
\tutenum\moscow            
\tutenum\naples            
\tutenum\cyprus            
\tutenum\nymegen           
\tutenum\caltech           
\tutenum\perugia           
\tutenum\peters            
\tutenum\cmu               
\tutenum\potenza           
\tutenum\prince            
\tutenum\riverside         
\tutenum\rome              
\tutenum\salerno           
\tutenum\ucsd              
\tutenum\sofia             
\tutenum\korea             
\tutenum\taiwan            
\tutenum\tsinghua          
\tutenum\purdue            
\tutenum\psinst            
\tutenum\zeuthen           
\tutenum\eth               

{
\parskip=0pt
\noindent
{\bf The L3 Collaboration:}
\ifx\selectfont\undefined
 \baselineskip=10.8pt
 \baselineskip\baselinestretch\baselineskip
 \normalbaselineskip\baselineskip
 \ixpt
\else
 \fontsize{9}{10.8pt}\selectfont
\fi
\medskip
\tolerance=10000
\hbadness=5000
\raggedright
\hsize=162truemm\hoffset=0mm
\def\r{\rlap,}
\noindent

P.Achard\r\tute\geneva\ 
O.Adriani\r\tute{\florence}\ 
M.Aguilar-Benitez\r\tute\madrid\ 
J.Alcaraz\r\tute{\madrid}\ 
G.Alemanni\r\tute\lausanne\
J.Allaby\r\tute\cern\
A.Aloisio\r\tute\naples\ 
M.G.Alviggi\r\tute\naples\
H.Anderhub\r\tute\eth\ 
V.P.Andreev\r\tute{\lsu,\peters}\
% ...
\newpage
\begin{list}{A}{\itemsep=0pt plus 0pt minus 0pt\parsep=0pt plus 0pt minus 0pt
                \topsep=0pt plus 0pt minus 0pt}
\item[\aachen]
 III. Physikalisches Institut, RWTH, D-52056 Aachen, Germany$^{\S}$
\item[\nikhef] National Institute for High Energy Physics, NIKHEF, 
     and University of Amsterdam, NL-1009 DB Amsterdam, The Netherlands
\item[\mich] University of Michigan, Ann Arbor, MI 48109, USA
\item[\lapp] Laboratoire d'Annecy-le-Vieux de Physique des Particules, 
     LAPP,IN2P3-CNRS, BP 110, F-74941 Annecy-le-Vieux CEDEX, France
\item[\basel] Institute of Physics, University of Basel, CH-4056 Basel,
     Switzerland
\item[\lsu] Louisiana State University, Baton Rouge, LA 70803, USA
\item[\beijing] Institute of High Energy Physics, IHEP, 
  100039 Beijing, China$^{\triangle}$ 
\item[\bologna] University of Bologna and INFN-Sezione di Bologna, 
     I-40126 Bologna, Italy
\item[\tata] Tata Institute of Fundamental Research, Mumbai (Bombay) 400 005, India
\item[\ne] Northeastern University, Boston, MA 02115, USA
\item[\bucharest] Institute of Atomic Physics and University of Bucharest,
     R-76900 Bucharest, Romania
\item[\budapest] Central Research Institute for Physics of the 
     Hungarian Academy of Sciences, H-1525 Budapest 114, Hungary$^{\ddag}$
\item[\mit] Massachusetts Institute of Technology, Cambridge, MA 02139, USA
\item[\panjab] Panjab University, Chandigarh 160 014, India
\item[\debrecen] KLTE-ATOMKI, H-4010 Debrecen, Hungary$^\P$
\item[\dublin] Department of Experimental Physics,
  University College Dublin, Belfield, Dublin 4, Ireland
\item[\florence] INFN Sezione di Firenze and University of Florence, 
     I-50125 Florence, Italy
\item[\cern] European Laboratory for Particle Physics, CERN, 
     CH-1211 Geneva 23, Switzerland
\item[\wl] World Laboratory, FBLJA  Project, CH-1211 Geneva 23, Switzerland
\item[\geneva] University of Geneva, CH-1211 Geneva 4, Switzerland
\item[\hamburg] University of Hamburg, D-22761 Hamburg, Germany
\item[\hefei] Chinese University of Science and Technology, USTC,
      Hefei, Anhui 230 029, China$^{\triangle}$
\item[\lausanne] University of Lausanne, CH-1015 Lausanne, Switzerland
\item[\lyon] Institut de Physique Nucl\'eaire de Lyon, 
     IN2P3-CNRS,Universit\'e Claude Bernard, 
     F-69622 Villeurbanne, France
\item[\madrid] Centro de Investigaciones Energ{\'e}ticas, 
     Medioambientales y Tecnol\'ogicas, CIEMAT, E-28040 Madrid,
     Spain${\flat}$ 
\item[\florida] Florida Institute of Technology, Melbourne, FL 32901, USA
\item[\milan] INFN-Sezione di Milano, I-20133 Milan, Italy
\item[\moscow] Institute of Theoretical and Experimental Physics, ITEP, 
     Moscow, Russia
\item[\naples] INFN-Sezione di Napoli and University of Naples, 
     I-80125 Naples, Italy
\item[\cyprus] Department of Physics, University of Cyprus,
     Nicosia, Cyprus
\item[\nymegen] Radboud University and NIKHEF, 
     NL-6525 ED Nijmegen, The Netherlands
\item[\caltech] California Institute of Technology, Pasadena, CA 91125, USA
\item[\perugia] INFN-Sezione di Perugia and Universit\`a Degli 
     Studi di Perugia, I-06100 Perugia, Italy   
\item[\peters] Nuclear Physics Institute, St. Petersburg, Russia
\item[\cmu] Carnegie Mellon University, Pittsburgh, PA 15213, USA
\item[\potenza] INFN-Sezione di Napoli and University of Potenza, 
     I-85100 Potenza, Italy
\item[\prince] Princeton University, Princeton, NJ 08544, USA
\item[\riverside] University of Californa, Riverside, CA 92521, USA
\item[\rome] INFN-Sezione di Roma and University of Rome, ``La Sapienza",
     I-00185 Rome, Italy
\item[\salerno] University and INFN, Salerno, I-84100 Salerno, Italy
\item[\ucsd] University of California, San Diego, CA 92093, USA
\item[\sofia] Bulgarian Academy of Sciences, Central Lab.~of 
     Mechatronics and Instrumentation, BU-1113 Sofia, Bulgaria
\item[\korea]  The Center for High Energy Physics, 
     Kyungpook National University, 702-701 Taegu, Republic of Korea
\item[\taiwan] National Central University, Chung-Li, Taiwan, China
\item[\tsinghua] Department of Physics, National Tsing Hua University,
      Taiwan, China
\item[\purdue] Purdue University, West Lafayette, IN 47907, USA
\item[\psinst] Paul Scherrer Institut, PSI, CH-5232 Villigen, Switzerland
\item[\zeuthen] DESY, D-15738 Zeuthen, Germany
\item[\eth] Eidgen\"ossische Technische Hochschule, ETH Z\"urich,
     CH-8093 Z\"urich, Switzerland
\item[\S]  Supported by the German Bundesministerium 
        f\"ur Bildung, Wissenschaft, Forschung und Technologie.
\item[\ddag] Supported by the Hungarian OTKA fund under contract
numbers T019181, F023259 and T037350.
\item[\P] Also supported by the Hungarian OTKA fund under contract
  number T026178.
\item[$\flat$] Supported also by the Comisi\'on Interministerial de Ciencia y 
        Tecnolog{\'\i}a.
\item[$\sharp$] Also supported by CONICET and Universidad Nacional de La Plata,
        CC 67, 1900 La Plata, Argentina.
\item[$\triangle$] Supported by the National Natural Science
  Foundation of China.
\end{list}
}
\vfill


\newpage

%
%

\begin{table}[h]
\begin{center}
\begin{tabular}{|c|c|c|c|c|c|c|c|c|}
      \hline
$\sqrt{s}$ (GeV) &   188.6 &  191.6 & 195.6 &  199.5 &  201.7 & $202.5-205.5$ &  $205.5-207.5$ &  $207.5-209.2$ \\   
$\cal{L}$ (pb$^{-1})$ & 176.8 &  29.7 & 83.9 &  82.8 &  39.1 &  77.8 &  131.4 & 8.2 \\ 
      \hline
\end{tabular}
\end{center}
        \icaption[]{\label{tab:data}
        Centre-of-mass energies and 
        corresponding integrated luminosities, $\cal{L}$, considered
        in this analysis.}
\end{table}

\begin{table} [h]
  \begin{center}
    \begin{tabular}{|c|c|c|c|}
      \cline{2-4}
      \multicolumn{1}{c|}{} &
      Preselection & Light-Higgs selection & Heavy-Higgs selection \\
      \hline
      Data            & 779             &  345            & 130             \\
      Standard Model  & $771.8 \pm 3.6$ &  $347.2\pm 2.0$ & $127.1\pm 1.8$  \\
      \hline
      Two-photon interactions    & $\phantom{00}6.4 \pm1.6 $& $-$                      & $\phantom{00}2.7 \pm1.1 $  \\ 
      Two-fermion final states   & $\phantom{0}69.9 \pm1.6 $& $\phantom{00}2.6 \pm0.3 $& $\phantom{0}21.4 \pm0.8 $  \\
      Four-fermion final states  & $          695.5 \pm2.8 $& $          344.6 \pm2.0 $& $          103.0 \pm1.1 $  \\
      \hline
    \end{tabular} 
    \icaption[]{\label{tab:hadrons} Results of the selection of events
    with hadrons and missing energy. The lower part of the table
    details the composition of the expected Standard Model sample. The
    uncertainties reflect the limited background Monte Carlo
    statistics.}
  \end{center}
\end{table}

\begin{table} [h]
  \begin{center}
    \begin{tabular}{|c|c|c|c|}
      \cline{2-4}
      \multicolumn{1}{c|}{} &
      \multicolumn{3}{c|}{Efficiency (\%)} \\
      \hline
      $m_{\rm h}$ (GeV) & $\rm Z \ra q\bar{q}$ & $\rm Z \ra \epem$ & $\rm Z \ra \mu^+ \mu^-$\\
      \hline
\phantom{0}60   &$ 49.0\pm 1.5 $ &$ 34.2\pm 0.9 $&$ 22.4\pm 0.8 $\\ 
\phantom{0}70   &$ 49.8\pm 1.6 $ &$ 38.0\pm 0.8 $&$ 26.6\pm 0.7 $\\ 
\phantom{0}80   &$ 49.1\pm 1.8 $ &$ 44.9\pm 0.8 $&$ 32.7\pm 0.8 $\\ 
\phantom{0}90   &$ 50.2\pm 1.9 $ &$ 49.9\pm 0.8 $&$ 31.2\pm 0.8 $\\ 
          100   &$ 49.4\pm 1.9 $ &$ 40.1\pm 0.8 $&$ 27.0\pm 0.7 $\\ 
          110   &$ 47.6\pm 1.7 $ &$ 23.3\pm 0.8 $&$ 14.8\pm 0.7 $\\ 
      \hline
    \end{tabular} 
    \icaption[]{\label{tab:effi} Selection efficiencies as a function of the mass of the invisibly-decaying
    Higgs boson. The uncertainties are due to the limited signal Monte
    Carlo statistics.}
  \end{center}
\end{table}

\begin{table} [h]
  \begin{center}
    \begin{tabular}{|c|c|c|c|c|}
      \cline{2-5}
      \multicolumn{1}{c|}{} &
      \multicolumn{2}{c|}{$\rm Z \ra \epem$}      &
      \multicolumn{2}{c|}{$\rm Z \ra \mu^+ \mu^-$}\\
      \cline{2-5}
      \multicolumn{1}{c|}{} &
      Preselection & Selection & Preselection & Selection \\
      \hline
      Data            & 147\phantom{.0} &  6\phantom{.0} & 115\phantom{.0}   & \phantom{0}9\phantom{.0}    \\
      Standard Model  & 136.4           &  9.7           & 130.2             & 11.1 \\
      \hline
      $\EE \to \EE (\gamma)$      & \phantom{0}19.7  & \phantom{0}0.3   & $-$            & $-$            \\ 
      $\EE \to \MM (\gamma)$      & $-$              & $-$              &           12.3 & \phantom{0}0.8 \\ 
      $\EE \to \TT (\gamma)$      & \phantom{00}1.5  & $-$              & \phantom{0}1.1 & $-$            \\ 
      Two-photon interactions     & \phantom{00}6.9  & $-$              &           30.6 & $-$            \\ 
      Four-fermion final states   & 108.3            & 9.4              &           86.2 & 10.3           \\
      \hline
    \end{tabular} 
    \icaption[]{\label{tab:emu}  Results of the selection of events
    with leptons and missing energy. The lower part of the table
    details the composition of the expected Standard Model sample. The statistical uncertainties on
    the background estimation are negligible.}
  \end{center}
\end{table}

\begin{table} [h]
  \begin{center}
    \begin{tabular}{|c|c|c|c|c|c|c|}
      \cline{2-7}
      \multicolumn{1}{c|}{} &
      \multicolumn{2}{c|}{$\rm Z \ra q\bar{q}$}   &
      \multicolumn{2}{c|}{$\rm Z \ra \epem$}      &
      \multicolumn{2}{c|}{$\rm Z \ra \mu^+ \mu^-$}\\
      \cline{2-7}
      \multicolumn{1}{c|}{} &
      Signal & Background & Signal & Background & Signal & Background \\
      \hline
      Monte Carlo statistics    & 1.4 \% & 5.7 \% & 1.9 \% &          $<0.1$ \% &  2.5 \% &          $<0.1$ \% \\ 
      Background cross sections & $-$    & 3.8 \% & $-$    & \phantom{0..}5.0  \% &  $-$    & \phantom{0..}5.0  \% \\
      Detector response         & 2.0 \% & 4.9 \% & 2.0 \% & \phantom{0..}2.7  \% &  2.5 \% & \phantom{0..}4.1  \% \\
      \hline									  			      	
      Total                     & 2.4 \% & 8.4 \% & 2.8 \% & \phantom{0..}5.7  \% &  3.5 \% & \phantom{0..}6.5  \% \\
      \hline
    \end{tabular} 
    \icaption[]{\label{tab:syst}Relative systematic uncertainties on the
      signal efficiency and background normalisation for each analysis channel.}
  \end{center}
\end{table}

\clearpage

%
%

\newpage
               
\newpage

\newpage

\newpage
                                
\newpage

\newpage

\end{document}